\documentclass[12pt]{article}
\def\baselinestretch{0.98}
\usepackage{cite}
\usepackage{multirow}
\usepackage{amssymb, amsmath, amsopn, amsthm}
\usepackage{bm,amsfonts,array,calc,rotating}
\usepackage{epsfig}
\usepackage{graphicx}
\usepackage{color}
\usepackage{epic}

\definecolor{orange}{cmyk}{0,0.5,1,0}

\newcommand{\ra}{\rightarrow}
\usepackage[T1]{fontenc}
\numberwithin{equation}{section}

\long\def\@makecaption#1#2{%
  \vskip\abovecaptionskip
  \sbox\@tempboxa{{\bf #1:} #2}%
  \ifdim \wd\@tempboxa >\hsize
    {\small\bf #1:} {\small #2}\par
  \else
    \global \@minipagefalse
    \hb@xt@\hsize{\hfil\box\@tempboxa\hfil}%
  \fi
  \vskip\belowcaptionskip}
\font\cmss=cmss10 \font\cmsss=cmss10 at 7pt
\def\IZ{\relax\ifmmode\mathchoice
{\hbox{\cmss Z\kern-.4em Z}}{\hbox{\cmss Z\kern-.4em Z}}
{\lower.9pt\hbox{\cmsss Z\kern-.4em Z}} {\lower1.2pt\hbox{\cmsss
Z\kern-.4em Z}}\else{\cmss Z\kern-.4em Z}\fi}

\def\sqr#1#2{{\vcenter{\vbox{\hrule height.#2pt
 \hbox{\vrule width.#2pt height#1pt \kern#1pt
 \vrule width.#2pt}\hrule height.#2pt}}}}

\begin{document}

\begin{flushright}
\baselineskip=12pt \normalsize
{MIFPA-10-22}\\
{TUW-10-06}\\
\smallskip
\end{flushright}

\begin{center}
\Large {\textbf{Flipped $SU(5)$ GUTs from $E_8$ Singularities in F-theory}} \\[2cm]
\normalsize Ching-Ming Chen$^{\sharp}$\footnote{\tt
cmchen@hep.itp.tuwien.ac.at} and Yu-Chieh
Chung$^{\natural}$\footnote{\tt ycchung@physics.tamu.edu}
\\[.25in]
\textit{$^{\sharp}$Institute for Theoretical Physics, Vienna
University of Technology\\
Wiedner Hauptstrasse 8-10, A-1040 Vienna, AUSTRIA} \\[0.25in]
\textit{$^{\natural}$Department of Physics $\&$ Astronomy, Texas A$\&$M University\\ College Station, TX 77843, USA} \\[2.5cm]
\end{center}

\renewcommand{\baselinestretch}{1.5}
\setlength{\baselineskip}{18pt}

\begin{abstract}
In this paper we construct supersymmetric flipped $SU(5)$ GUTs
from $E_8$ singularities in F-theory. We start from an $SO(10)$
singularity unfolded from an $E_8$ singularity by using an $SU(4)$
spectral cover. To obtain realistic models, we consider $(3,1)$
and $(2,2)$ factorizations of the $SU(4)$ cover. After turning on
the massless $U(1)_X$ gauge flux, we obtain the $SU(5)\times
U(1)_X$ gauge group. Based on the well-studied geometric
backgrounds in the literature, we demonstrate several models and
discuss their phenomenology.

\end{abstract}
\vspace{3cm}

%\today

\newpage
\setcounter{page}{1}

\setcounter{footnote}{0}

\pagenumbering{arabic}

\pagestyle{plain}

\section{Introduction}

String theory is a ten-dimensional theory of quantum gravity and
so far is the most promising candidate for a fundamental unified
theory. To build connections to the physics at a low energy scale,
string theorists have been using the techniques of
compactification to construct models in four-dimensional
spacetime. F-theory \cite{Vafa:1996zn,Vafa:1996yn,Vafa:1996xn}(see
\cite{Denef:2008wq} for review) is a twelve-dimensional geometric
extension of string theory where one can engineer gauge theories
from a geometric approach \cite{Bershadsky:1996nh, Katz:1996xe}.
We are interested in how gauge theories realized by F-theory can
accommodate Grand Unified Theory (GUT) models. Recently, extensive
studies of GUT local models and their corresponding phenomenology
in F-theory have been undertaken in \cite{King:2010mq,
Chung:2009ib, Vafa:NoncommutativeandYukawa, Conlon:local02,
Conlon:local01, Randall:localFlavor, Heckman:local02a, Li:local03,
Ibanez:localFlavor02, Ibanez:locaFlavorl01, BHV:2008I, BHV:2008II,
Donagi:2008sl,Donagi:2008ll, Heckman:local01, Heckman:local02,
Heckman:local03, Heckman:localFlavor01, Heckman:localFlavor02,
Heckman:localFlavor03, Nanopoulos:local01, Nanopoulos:local02,
Blumenhagen:local, Chen:2009me, Bourjaily:local01,
Bourjaily:local02}. In addition, supersymmetry breaking has been
discussed in \cite{Blumenhagen:SUSYbreaking,
Buchbinder:SUSYbreaking, Caltech:SUSYbreaking03,
Caltech:SUSYbreaking02, Caltech:SUSYbreaking01}, and the
application to cosmology has been studied in
\cite{Heckman:cosmology}.  Semi-local and global model building in
F-theory were particularly discussed in \cite{Choi:2010su,
Dudas:2009hu, Hayashi:2010zp, Heckman:2010pv, Cvetic:2009ah,
Cvetic:2010rq, Blumenhagen:2010ja, Jockers:2009ti, Grimm:2009yu,
Grimm:2009sy, Esole:global01, Curio:global, Collinucci:global03,
Collinucci:global02, Watari:global, Hebecker:global,
Cordova:Decouplinggeom, Donagi:global, Caltech:global01,
Caltech:global02, Caltech:global03, Blumenhagen:global01,
Blumenhagen:global02, Tartar:globalFlavor01,
Tartar:globalFlavor02, Other:global,other:global02,
Grimm:global01}. Systematic studies of how models of higher rank
GUT groups, such as $SO(10)$, are embedded into the compact
geometry in F-theory have not been fully investigated. To this
end, we are interested in the $SO(10)$ subgroup $SU(5)\times
U(1)_X$ which is realized as the flipped $SU(5)$ GUT
\cite{dimitri, smbarr, AEHN-0}. Although local flipped $SU(5)$
models have been discussed in F-theory, we study the model as a
semi-local construction. In this paper we shall build flipped
$SU(5)$ models by unfolding an $E_8$ singularity via the $SO(10)$
gauge group.

To construct flipped $SU(5)$ models in the four-dimensional
spacetime, we compactify F-theory on an elliptically fibered
Calabi-Yau fourfold $X_4$ with a base threefold $B_3$. We adopt a
bottom-up approach to construct models in the decoupling limit to
avoid full F-theory on a complicated elliptically fibered
Calabi-Yau fourfold. More precisely, we consider a contractible
complex surface $S$ inside $B_3$ such that we can reduce full
F-theory on $X_4$ to an effective eight-dimensional supersymmetric
gauge theory on $\mathbb{R}^{3,1}\times S$. In this paper the
surface $S$ is assumed to be a del Pezzo surface
\cite{delPezzo:01, delPezzo:02}. Since we will construct flipped
$SU(5)$ models from an $SO(10)$ gauge group, we have to engineer
the singularities of types $D_5$, $D_6$, $E_6$, and $E_7$ in the
Calabi-Yau fourfold $X_4$. Because these singularities can be
embedded into a single singularity $E_8$, we start our discussion
from the $E_8$ singularity and unfold it into a $D_5$ singularity.

Generally, one may turn on certain fluxes to obtain the chiral
spectrum. In F-theory, there is a four-form $G$-flux, which
consists of three-form fluxes and gauge fluxes. In type IIB
theory, these three-form fluxes produce a back-reaction in the
background geometry. It has been shown in
\cite{Vafa:NoncommutativeandYukawa,Marchesano:2009rz} that the
three-form fluxes induce non-commutative geometric structures and
also modify the texture of the Yukawa couplings. F-theory in Fuzzy
space also has been studied in \cite{Heckman:2010pv}. In this
paper we shall turn off these three-form fluxes and focus only on
the gauge fluxes. The gauge $U(1)_X$ flux is able to break the
gauge group $SO(10)$ down to $SU(5)\times U(1)_X$. It was shown in
\cite{Donagi:global, Donagi:2008ll} that the spectral cover
construction naturally encodes the unfolding information of an
$E_8$ singularity as well as the gauge fluxes. In this paper we
shall focus on the $SU(4)$ spectral cover encoding the $SO(10)$
singularity from unfolding $E_8$. The four-dimensional low-energy
spectrum of the flipped $SU(5)$ model is then determined by the
cover fluxes and the $U(1)_X$ flux.

The $SU(4)$ spectral cover has many interesting properties. From
the subgroup decomposition of $E_8$, one can find that there is no
explicit presentation of $\overline{\bf 10}$. In addition, the
cover associated to the ${\bf 10}$ representation forms a
double-curve and along this curve there are co-dimension two
singularities. After resolving the singularities along the curve,
one finds that the net chirality of the $\bf 10$ curve vanishes
\cite{Other:global}. Since the background geometry generically
determines the $G$ flux, there are not many degrees of freedom
left to adjust the chirality on the $\bf 16$ curve to create
three-generation models. These ideas motivate us to consider
factorizing the spectral cover \cite{Caltech:global02,
Caltech:global03, Blumenhagen:global01, Blumenhagen:global02,
Grimm:global01} to introduce additional parameters for model
building. We consider two possibilities of splitting the $SU(4)$
spectral cover: (3,1) and (2,2) factorizations. The curve of the
fundamental representation is then divided into two $\bf 16$
curves, while generically the $\bf 10$ curve is detached into
three. However, due to the monodromy structure there are only two
$\bf 10$ curves in the (3,1) case.

In semi-local $SO(10)$ GUTs, there exists only the $\bf 16\, 16\,
10$ Yukawa coupling from the enhancement to an $E_7$ singularity.
The GUT Higgs fields coming from the adjoints or other
representations such as $\bf 45$, $\bf 54$, or $\bf 120$ are
absent in the F-theory construction. Therefore, the most
convincing way to break the $SO(10)$ gauge group is turning on the
$U(1)_X$ flux on the GUT surface $S$. This $U(1)_X$ gauge field
can be massless \cite{BHV:2008I, Donagi:2008sl, Verlinde:2005jr},
so we can interpret the gauge group as the flipped $SU(5)$ model
after turning on such a flux. With non-trivial restrictions to the
curves, this $U(1)_X$ flux generically modifies the net chirality
of matter localized on these curves. We may identify the flipped
$SU(5)$ superheavy Higgs fields with one of the ${\bf
10}+\overline{\bf 10}$ vector-like pairs in the spectrum for
further gauge breaking to MSSM.

The organization of the rest of the paper is as follows: in
section 2, we briefly review the local geometry of an elliptically
fibered Calabi-Yau fourfold with $ADE$ singularities and the
$SU(4)$ spectral cover. In section 3, we study $(3,1)$ and $(2,2)$
factorizations of the $SU(4)$ cover. In section 4, we construct
cover fluxes and compute the chirality of matter localized on each
curve for the $(3,1)$ and $(2,2)$ cover factorizations. In section
5, we briefly review the $D3$ tadpole cancellation in F-theory. We
also give explicit formulae of geometric and cover flux
contributions in the tadpole cancellation. In section 6, we
demonstrate several examples of flipped $SU(5)$ models and discuss
their phenomenology. We summarize and conclude in section 7.

\section{Preliminaries}

\subsection{Elliptically fibered Calabi-Yau Fourfolds and $ADE$ Singularities}

Let us consider an elliptically fibered Calabi-Yau fourfold
$\pi:X_4\ra B_3$ with a section $\sigma_{B_3}: B_3\ra X_4$. Due to
the presence of the section $\sigma_{B_3}$, $X_4$ can be described
by the Weierstrass form:
\begin{equation}
y^2=x^3+fx+g,\label{Weierstrass model}
\end{equation}
where $f$ and $g$ are sections of suitable line bundles over
$B_3$. More precisely, to maintain Calabi-Yau condition
$c_1(X_4)=0$, it is required that\footnote{The symbol $\Gamma(L)$
stands for a set of global sections of the bundle $L$.}
$f\in\Gamma (K^{-4}_{B_3})$ and $g\in\Gamma (K^{-6}_{B_3})$, where
$K_{B_3}$ is the canonical bundle of $B_3$. Let $\Delta\equiv
4f^3+27g^2$ be the discriminant of the elliptic fibration Eq.
(\ref{Weierstrass model}) and $S$ be one component of the locus
$\{\Delta=0\}$ where elliptic fibers degenerate. In the vicinity
of $S$, one can regard $X_4$ as an ALE fibration over the surface
$S$. To construct $SO(10)$ and flipped $SU(5)$ GUT models, one can
start with engineering a $D_5$ singularity corresponding to the
gauge group $SO(10)$ in the following way. Let $z$ be a section of
the normal bundle $N_{S/B_3}$ of $S$ in $B_3$ and the zero section
then represents the surface $S$. Since $f$ and $g$ are sections of
some line bundles over $B_3$, one can locally expand $f$ and $g$
in terms of $z$ as follows:
\begin{equation}
f=3\sum_{k=0}^4f_k(u,v)z^k,\;\;\;\;\;g=2\sum_{l=0}^6g_l(u,v)z^l,
\end{equation}
where $(u,v)$ are coordinates of $S$ and the prefactors $2$ and
$3$ are just for convenience. Then the Weierstrass form Eq.
(\ref{Weierstrass model}),
\begin{equation}
y^2=x^3+3\sum_{k=0}^4f_k(u,v)z^kx+2\sum_{l=0}^6g_l(u,v)z^l,\label{ALE
model}
\end{equation}
describes an ALE fibration over $S$, where
$f_k\in\Gamma(K_{B_3}^{-4}\otimes \mathcal{O}_{B_3}(-kS))$ and
$g_l\in\Gamma(K_{B_3}^{-6}\otimes
\mathcal{O}_{B_3}(-lS))$.\footnote{By adjunction formula,
$K_S=K_{B_3}\otimes N_{S/B_3}|_{S}$, we have
$f_k\in\Gamma(K_{S}^{-4}\otimes N^{4-k}_{S/B_3})$ and
$g_l\in\Gamma(K_{S}^{-6}\otimes N^{6-l}_{S/B_3})$, where $K_S$ is
the canonical bundle of $S$. } According to the Kodaira
classification of singular elliptic fibers, one can classify the
singularity of an elliptic fibration by the vanishing order of
$f$, $g$, and $\Delta$, denoted by ${\rm ord}(f)$, ${\rm ord}(g)$,
and ${\rm ord}(\Delta)$, respectively. We summarize the relevant
$ADE$ classification and corresponding gauge groups in Table
{\ref{Kodaira's Classification}}. A detailed list can be found in
\cite{Donagi:2008ll}.
\begin{table}[h]
\begin{center}
\renewcommand{\arraystretch}{0.8}
\begin{tabular}{|c|c|c|c|c|c|c|c|} \hline

Singularity &  ${\rm ord}(f)$ & ${\rm ord}(g)$ & ${\rm ord}(\Delta)$ & Gauge Group \\
\hline
 $A_n$& 0& 0& $n+1$ & $SU(n+1)$\\
\hline $D_{n+4}$ & $\geqslant 2$ &  3 & $n+6$
 & $SO(2n+8)$ \\
\hline $D_{n+4}$ & 2& $\geqslant 3$ & $n+6$
 & $SO(2n+8)$ \\
\hline $E_6$ & $\geqslant 3$& 4& 8
& $E_6$\\
\hline $E_7$ & 3 & $\geqslant 5$ & 9
 & $E_7$\\
\hline $E_8$ & $\geqslant 4$ & 5 & 10
 & $E_8$\\
\hline
\end{tabular}
\caption{$ADE$ singularities and corresponding gauge
groups.}\label{Kodaira's Classification}
\end{center}
\end{table}
According to Table {\ref{Kodaira's Classification}}, a $D_5$
singularity corresponds to the case of $({\rm ord}(f),{\rm
ord}(g),{\rm ord}(\Delta))=(\geqslant 2,3,7)$ or $(2,\geqslant
3,7)$. Recall that $S$ is the locus $\{z=0\}$. To obtain a $D_5$
singularity, the vanishing orders of $f$ and $g$ at $z=0$ are
required to be two and three, respectively\footnote{One can show
that in this case the only consistent triplet vanishing orders for
a $D_5$ singularity is $({\rm ord}(f),{\rm ord}(g),{\rm
ord}(\Delta))=(2,3,7)$. The higher order terms are irrelevant to
the singularity. However, they may change the monodromy group
\cite{Hayashi:2010zp}.}. Let us consider the sections $f$ and $g$
to be
\begin{equation}
f=3(f_2z^2+f_3z^3),\;\;\;\;\;\;g=2(g_3z^3+g_4z^4+g_5z^5)\label{fg}.
\end{equation}
Then the corresponding discriminant is given by
\begin{eqnarray}
\Delta &=&
cz^6[(f^3_2+g^2_3)+(3f^2_{2}f_3+2g_3g_4)z+(3f_2f^2_3+g^2_4+2g_3g_5)z^2\nonumber
\\&+&(f_3^3+2g_4g_5)z^3+\mathcal{O}(z^4)],
\end{eqnarray}
where $c=4\cdot 27$. To obtain ${\rm ord}(\Delta)=7$, let us set
$f_2=-h^2$ and $g_3=h^3$, where
$h\in\Gamma(K^{-2}_{B_3}\otimes\mathcal{O}_{B_3}(-S))$. Then the
discriminant is reduced to
\begin{equation}
\Delta =cz^7[(3h^4f_3+2h^3g_4)+(-3h^2f^2_3+g^2_4+2h^3g_5)z+
(f_3^3+2g_4g_5)z^2+\mathcal{O}(z^3)]. \label{Delta_D5}
\end{equation}
The singularity of ALE fibration is now characterized by the
sections $\{h,f_3,g_4,g_5\}$. When $h=0$, one can find that $({\rm
ord}(f),{\rm ord}(g),{\rm ord}(\Delta))=(3,4,8)$ at the locus
$\{z=0\}\cap\{h=0\}$. It follows from the Kodaira classification
that the singularity is enhanced to $E_6$. When $3hf_3+2g_4=0$,
the triplet vanishing orders becomes $(2,3,8)$, which implies that
the singularity at the locus $\{z=0\}\cap\{3hf_3+2g_4=0\}$ is
$D_6$ and that the corresponding enhanced gauge group is $SO(12)$.
In a similar manner, one can find the codimension two
singularities corresponding to $E_7$ and $SO(14)$ in $S$. We
summarize the results in Table \ref{Singularity for Weierstrass
form}.
\begin{table}[h]
\begin{center}
\renewcommand{\arraystretch}{1}
\begin{tabular}{|c|c|c|c|c|c|c|c|} \hline
Gauge Group & $({\rm ord}(f),{\rm ord}(g),{\rm ord}(\Delta))$ & Locus\\
\hline
$SO(10)$& $(2,3,7)$ & $\{z=0\}$\\
\hline
$E_6$ & $(3,4,8)$ & $\{z=0\}\cap\{h=0\}$\\
\hline
$SO(12)$ & $(2,3,8)$ & $\{z=0\}\cap\{3hf_3+2g_4=0\}$\\
\hline
$E_7$ & $(3,5,9)$ & $\{z=0\}\cap\{h=0\}\cap\{g_4=0\}$\\
\hline
$SO(14)$ & $(2,3,9)$ & $\{z=0\}\cap\{3hf_3+2g_4=0\}\cap\{3f_3^2-8hg_5=0\}$\\
\hline
\end{tabular}
\caption{Gauge enhancements and corresponding loci.}
\label{Singularity for Weierstrass form}
\end{center}
\end{table}

For later use, it is convenient to introduce the Tate form of the
fibration:
\begin{equation}
y^2=x^3+\mathbf{b}_4x^2z+\mathbf{b}_3yz^2+\mathbf{b}_2xz^3+\mathbf{b}_0z^5,\label{Tate
form}
\end{equation}
where $\mathbf{b}_m\in \Gamma(K_S^{m-6}\otimes N_{S/B_3})$.
Actually, Eq.~(\ref{Tate form}) is nothing more than the unfolding
of an $E_8$ singularity to a singularity of $SO(10)$. Notice that
by comparing Eq.~(\ref{Tate form}) with Eqs.~(\ref{ALE model}) and
(\ref{fg}), one can obtain the relations between
$\{f_2,f_3,g_3,g_4,g_5\}$ and
$\{\mathbf{b}_0,\mathbf{b}_2,\mathbf{b}_3,\mathbf{b}_4\}$ as
follows:
\begin{equation}
\left\{\begin{array}{l} f_2=-\frac{1}{9}\mathbf{b}_4^2\\
f_3=\frac{1}{3}\mathbf{b}_2\\
g_3=\frac{1}{27}\mathbf{b}_4^3 \\
g_4=\frac{1}{8}\mathbf{b}_3^2 - \frac{1}{6}\mathbf{b}_2\mathbf{b}_4 \\
g_5=\frac{1}{2}\mathbf{b}_0.
\end{array}\label{mapping}   \right.
\end{equation}
With the relations in Eq.~(\ref{mapping}), the discriminant
Eq.~(\ref{Delta_D5}) becomes
\begin{eqnarray}
\Delta &=&\widetilde{c}z^7 \{16 \mathbf{b}_3^2 \mathbf{b}_4^3 +[
27\mathbf{b}_3^4 -72\mathbf{b}_2\mathbf{b}_3^2\mathbf{b}_4 -16\mathbf{b}_4^2(\mathbf{b}_2^2-4\mathbf{b}_0\mathbf{b}_4) ]z\nonumber \\
&+& [16\mathbf{b}_2(4\mathbf{b}_2^2 -18\mathbf{b}_0\mathbf{b}_4)+
216\mathbf{b}_0 \mathbf{b}_3^2]z^2+ \mathcal{O}(z^3) \},
\label{Delta_D5 01}
\end{eqnarray}
where $\widetilde{c}=\frac{1}{16}$. It follows from
Eq.~(\ref{mapping}) that the codimension one loci
$\{z=0\}\cap\{h=0\}$ and $\{z=0\}\cap\{3hf_3+2g_4\}$ in $S$ can be
equivalently expressed as $\{z=0\}\cap\{\mathbf{b}_4=0\}$ and
$\{z=0\}\cap\{\mathbf{b}_3=0\}$, respectively. Due to the gauge
enhancements, matter ${\bf 16}$ and ${\bf 10}$ are localized at
the loci of $E_6$ and $SO(12)$ singularities, respectively. One
can also find that the loci of codimension two singularities $E_7$
and $SO(14)$ in $S$ are
$\{z=0\}\cap\{\mathbf{b}_3=0\}\cap\{\mathbf{b}_4=0\}$ and
$\{z=0\}\cap\{\mathbf{b}_3=0\}\cap\{\mathbf{b}_2^2-4\mathbf{b}_0\mathbf{b}_4=0\}$,
respectively. At these loci, the corresponding gauge groups are
enhanced to $E_7$ and $SO(14)$, respectively\footnote{One can also
use Tate's algorithm to determine the singularity type of the Tate
form Eq. (\ref{Tate form}) \cite{Bershadsky:1996nh}.}. In
particular, the Yukawa coupling ${\bf 16\,16\,10}$ can be realized
at the points with $E_7$ singularities. We summarize the results
in Table \ref{Singularity for Tate form}.
\begin{table}[h]
\begin{center}
\renewcommand{\arraystretch}{1}
\begin{tabular}{|c|c|c|c|c|c|c|c|} \hline
Gauge Group & Locus & Object\\
\hline
$SO(10)$ & $\{z=0\}$ & GUT Seven-branes \\
\hline
$E_6$ &  $\{z=0\}\cap\{\mathbf{b}_4=0\}$ & Matter $\bf 16$\\
\hline
$SO(12)$ & $\{z=0\}\cap\{\mathbf{b}_3=0\}$ & Matter ${\bf 10}$\\
\hline
$E_7$ & $\{z=0\}\cap\{\mathbf{b}_3=0\}\cap\{\mathbf{b}_4=0\}$ & Yukawa Coupling ${\bf 16\,16\,10}$\\
\hline
$SO(14)$ & $\{z=0\}\cap\{\mathbf{b}_3=0\}\cap\{\mathbf{b}_2^2-4\mathbf{b}_0\mathbf{b}_4=0\}$ & Extra Coupling\\
\hline
\end{tabular}
\caption{Gauge enhancements in $SO(10)$ GUT geometry.}
\label{Singularity for Tate form}
\end{center}
\end{table}

\subsection{$SU(4)$ Spectral Cover}

\begin{figure}[h]
\center \large
\begin{picture}(400,80)
\thicklines
\multiput(60, 20)(40, 0){8}{\circle{10}} %
\put(260, 60){\circle{10}} \put(260, 25){\line(0,1){30}}%
\multiput(105, 20)(40, 0){6}{\line(1, 0){30}} %
\dashline[4]{6}(65,20)(95,20) %
\put(58,30){\small 1} \put(98,30){\small 2} \put(138,30){\small 3}
\put(178,30){\small 4} \put(218,30){\small 5} \put(263,30){\small
6} \put(298,30){\small 4} \put(338,30){\small 2}
\put(269,56){\small 3} %
\put(52,4){\small $\alpha_{-\theta}$} \put(97,4){\small
$\alpha_1$} \put(137,4){\small $\alpha_2$} \put(177,4){\small
$\alpha_3$} \put(217,4){\small $\alpha_4$} \put(257,4){\small
$\alpha_5$} \put(297,4){\small $\alpha_6$} \put(337,4){\small
$\alpha_7$} \put(240,57){\small $\alpha_8$}
\end{picture}
\caption{The extended $E_8$ Dynkin diagram and
indices}\label{Dynkin}
\end{figure}
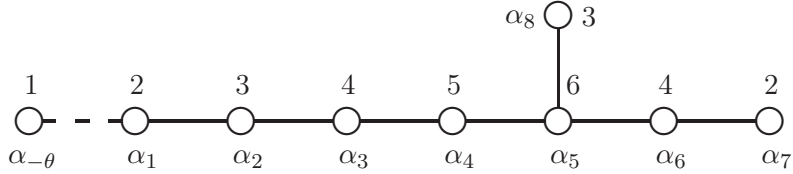

To engineer the $SO(10)$ gauge group from an $E_8$ singularity,
let us consider the following decomposition
\begin{equation}
\begin{array}{c@{}c@{}l@{}c@{}l}
E_8 &~\rightarrow~& SO(10)\times SU(4)_{\bot}\\
{\bf 248} &~\rightarrow~& ({\bf 1},{\bf 15})+({\bf 45},{\bf 1})+({\bf 10},{\bf 6})+({\bf 16},{\bf 4})+({\bf\overline{16}},{\bf\bar 4}).\\
\label{E_8 decomposition}
\end{array}
\end{equation}
and the Tate form of the fibration,
\begin{equation}
y^2=x^3+\mathbf{b}_4x^2z+\mathbf{b}_3yz^2+\mathbf{b}_2xz^3+\mathbf{b}_0z^5.\label{Tate
form 3}
\end{equation}
For simplicity, let us define $c_1\equiv c_1(S)$ and $t\equiv
-c_1(N_{S/B_{3}})$, then the homological classes of the sections
$x$, $y$, $z$, and $b_m$ can be expressed as
\begin{equation}
[x]=3(c_1-t),\;[y]=2(c_1-t),\;[z]=-t,\; [\mathbf{b}_m]=
(6-m)c_1-t\equiv\eta-mc_1.
\end{equation}
Recall that locally $X_4$ can be described by an ALE fibration
over $S$. Pick a point $p\in S$ and the fiber is an ALE space
denoted by ${\rm ALE}_p$. One can construct an ALE space by
resolving an orbifold $\mathbb{C}^2/\Gamma_{ADE}$, where
$\Gamma_{ADE}$ is a discrete subgroup of $SU(2)$
\cite{Douglas:1996sw}, for more information, see \cite{Math3,
Math2, Math1, Math4, Math5}. It was shown that the intersection
matrix of the exceptional 2-cycles corresponds to the Cartan
matrix of $ADE$ types. In this paper we will focus on engineering
the $SO(10)$ gauge group by unfolding an $E_8$ singularity. To
this end, let us consider $\alpha_i\in H_{2}({\rm
ALE}_p,\mathbb{Z}),\: i=1,2,...,8$ to be the roots\footnote{By
abuse of notation, the corresponding exceptional 2-cycles are also
denoted by $\alpha_i$} of $E_8$. The extended $E_8$ Dynkin diagram
with roots and Dynkin indices are shown in Fig \ref{Dynkin}.
Notice that $\alpha_{-\theta}$ is the highest root and satisfies
the condition
$\alpha_{-\theta}+2\alpha_1+3\alpha_2+4\alpha_3+5\alpha_4+6\alpha_5+4\alpha_6+2\alpha_7+3\alpha_8=0$.
To obtain $SO(10)$, we keep the volume of the cycles
$\{\alpha_4,\alpha_5,...,\alpha_8\}$ vanishing and then
$SU(4)_{\bot}$ is generated by $\{\alpha_1,\alpha_2,\alpha_3\}$.
An enhancement to $E_6$ happens when $\alpha_3$ or any of its
image under the Weyl permutation shrinks to zero size. Let
$\{\lambda_1,...,\lambda_4\}$ be the periods %\footnote{The period of a 2-cycle $\alpha$ is denoted by $\pi(\alpha)$.}
of these 2-cycles.
%\begin{equation}
%\left\{\begin{array}{l} \lambda_1=\pi(\alpha_3)\\
%\lambda_2=\pi(\alpha_3+\alpha_2)\\
%\lambda_3=\pi(\alpha_3+\alpha_2+\alpha_1)\\
%\lambda_4=\pi(\alpha_3+\alpha_2+\alpha_1+\alpha_{-\theta}).
%\end{array}\label{Lambda}   \right.
%\end{equation}
As described in \cite{Donagi:2008sl, Donagi:global}, the
information of theses $\lambda_i$ can be encoded in the
coefficients $\mathbf{b}_m$ in Eq.~(\ref{Tate form 3}) via the
following relations:
\begin{equation}
\left\{\begin{array}{l}\displaystyle \sum_{i}\lambda_i=\frac{b_1}{b_0}=0\\
\displaystyle\sum_{i<j}\lambda_i\lambda_j=\frac{b_2}{b_0}\\
\displaystyle\sum_{i<j<k}\lambda_i\lambda_j\lambda_k=\frac{b_3}{b_0}\\
\displaystyle\prod_{l}\lambda_l=\frac{b_4}{b_0},
\end{array}\label{Lambda}   \right.
\end{equation}
where $b_m\equiv \mathbf{b}_m|_{z=0}$. Equivalently,
$\{\lambda_1,...,\lambda_4\}$ can be regarded as the roots of the
equation
\begin{equation}
b_0\prod_{k} (s+\lambda_k)=b_0s^4+b_2s^2+b_3s+b_4=0. \label{affine
SU(4) cover}
\end{equation}
When $p\in S$ varies along $S$, Eq. (\ref{affine SU(4) cover})
defines a fourfold cover over $S$, called the fundamental $SU(4)$
spectral cover. This cover is a section of the canonical bundle
$K_S\ra S$. When $\lambda_i$ vanish, $\prod_{i}\lambda_i=b_4=0$ in
which the gauge group is enhanced to $E_6$ and matter ${\bf 16}$
is localized. According to the decomposition (\ref{E_8
decomposition}), matter ${\bf 10}$ corresponds to the
anti-symmetric representation ${\bf 6}$ of $SU(4)_{\bot}$,
associated to a sixfold cover $\mathcal{C}_{\wedge^2 V}^{(6)}$
over $S$. This associated cover $\mathcal{C}^{(6)}_{\wedge^2 V}$
can be constructed as follows:
\begin{equation}
b_0^2\prod_{i<j} (s+\lambda_i +\lambda_j)=b_0^2 s^6+ 2b_0 b_2 s^4
+( b_2^2 -4b_0 b_4)s^2 - b_3^2=0.\label{antisymmetry cover}
\end{equation}
Since matter ${\bf 10}$ corresponds to
$\lambda_i+\lambda_j=0,\;i\neq j$, it follows from Eq.
(\ref{antisymmetry cover}) that $b_3=0$, which means that matter
${\bf 10}$ is localized at the locus $\{b_3=0\}$ as shown in Table
\ref{Singularity for Tate form}. It is not difficult to see that
the spectral covers indeed encode the information of singularities
and gauge group enhancements. However, the spectral cover is even
more powerful. With it, we can construct a Higgs bundle to
calculate the chirality of matter ${\bf 16}$ and ${\bf 10}$ by
switching on a line bundle on the cover.

Let us define $X$ to be the total space of the canonical bundle
$K_S$ over $S$. Note that $X$ is a local Calabi-Yau threefold.
However, $X$ is non-compact%\footnote{The intersection of divisors
%may not be well-defined in a non-compact manifold.}
. To obtain a compact space, one can compactify $X$ to the total
space $\bar X$ of the projective bundle over $S$, $i.e.$
\begin{equation}
\bar{X}= {\mathbb{P}}(\mathcal{O}_S\oplus K_S),
\end{equation}
with a map $\pi:{\bar X}\ra S$, where $\mathcal{O}_S$ is the
trivial bundle over $S$. Notice that $\bar X$ is compact but no
longer a Calabi-Yau threefold. Let $\mathcal{O}(1)$ be a
hyperplane section of ${\mathbb{P}}^1$ fiber and denote its first
Chern class by $\sigma_{\infty}$. We define the homogeneous
coordinates of the fiber by $[U:V]$. Note that $\{U=0\}$ and
$\{V=0\}$ are sections of $\mathcal{O}(1)\otimes K_S$ and
$\mathcal{O}(1)$, while the class of $\{U=0\}$ and $\{V=0\}$ are
$\sigma\equiv\sigma_{\infty}-\pi^{\ast}c_1(S)$ and
$\sigma_{\infty}$, respectively. The intersection of $\{U=0\}$ and
$\{V=0\}$ is empty. Thus, one can obtain
$\sigma\cdot\sigma=-\sigma\cdot\pi^{\ast}c_1$. The affine
coordinate $s$ is defined by $s=U/V$. In $\bar X$, the $SU(4)$
cover Eq.~(\ref{affine SU(4) cover}) is homogenized as
\begin{equation}
\mathcal{C}_V^{(4)}:\;\;\;b_0U^4+b_2U^2V^2+b_3UV^3+b_4V^4=0\label{homo
SU(4) cover}
\end{equation}
with induced map $p_4: \mathcal{C}_V^{(4)}\ra S$. It is not
difficult to see that the homological class
$[\mathcal{C}^{(4)}_{V}]$ of the cover $\mathcal{C}^{(4)}_{V}$ is
given by $[\mathcal{C}^{(4)}_{V}]=4\sigma+\pi^{\ast}\eta$. One can
calculate the locus of the matter ${\bf 16}$ curve by intersection
of $[\mathcal{C}^{(4)}_{V}]$ with $\sigma$
\begin{equation}
[\mathcal{C}^{(4)}_{V}]\cap\sigma=(4\sigma+\pi^{\ast}\eta)\cdot\sigma=\sigma\cdot\pi^{\ast}(\eta-4c_1),
\end{equation}
which implies that $[\Sigma_{\bf 16}]=\eta-4c_1$ in $S$.
Alternatively, one could deduce this from the fact that the locus
of $\Sigma_{\bf 16}$ in $S$ is $\{b_4=0\}$. It follows from Eq.
(\ref{antisymmetry cover}) that the homological class of the cover
$\mathcal{C}^{(6)}_{\wedge^2V}$ is given by
\begin{equation}
[\mathcal{C}^{(6)}_{\wedge^2V}] =6\sigma+2\pi^{\ast}\eta
\end{equation}
Notice that $\mathcal{C}^{(6)}_{\wedge^2 V}$ is generically
singular. To solve this problem, one can consider intersection
$\tau \mathcal{C}_V \cap \mathcal{C}_V$ and define
\cite{Donagi:2004ia, other:global02}
\begin{eqnarray}
[D]=[\mathcal{C}^{(4)}_V]\cap [\mathcal{C}^{(4)}_V] -
[\mathcal{C}^{(4)}_V]\cap \sigma - [\mathcal{C}^{(4)}_V]\cap
3\sigma_{\infty}
\end{eqnarray}
where $\tau$ is a $\mathbb{Z}_2$ involution $V\ra -V$ acting on
the spectral cover\footnote{Note that there are double points on
$\Sigma_{\bf 10}$. One can resolve these double points by
blowing-up and then obtain resolved $\tilde{\Sigma}_{\bf 10}$ with
a mapping $\tilde{\pi}_D: D \rightarrow \tilde{\Sigma}_{\bf 10}$
of degree 4 and $[\tilde{\Sigma}_{\bf 10}]=\eta-3c_1$
\cite{Other:global}.}. The {\bf 10} curve can then be evaluated by
\begin{eqnarray}
[D]|_{\sigma}=4(\eta-3c_1),
\end{eqnarray}
which implies that $[\Sigma_{\bf 10}]=2\eta-6c_1$ in $S$.

To obtain chiral spectrum, we turn on a spectral line bundle
$\mathcal{L}$ on the cover $\mathcal{C}_{V}^{(4)}$. The
corresponding Higgs bundle is given by $V=p_{4\ast}\mathcal{L}$.
For an $SU(n)$ bundle, it is required that $c_1(V)=0$. It follows
that
\begin{equation}
c_1(p_{4\ast}\mathcal{L})=p_{4\ast}c_1(\mathcal{L})-\frac{1}{2}p_{4\ast}r,
\end{equation}
where $r$ is the ramification divisor given by
$r=p_{4\ast}c_1-c_1(\mathcal{C}_{V}^{(4)})$. It is convenient to
define the cover flux $\gamma$ by
\begin{equation}
c_1(\mathcal{L})=\lambda\gamma+\frac{1}{2}r,
\end{equation}
where $\lambda$ is a parameter used to compensate the non-integral
class $\frac{1}{2}r$. The traceless condition
$c_1(p_{4\ast}\mathcal{L})=0$ is then equivalent to the condition
$p_{4\ast}\gamma=0$. One can show that
\begin{equation}
\gamma=(4-p^{\ast}_4p_{4\ast})(\mathcal{C}^{(4)}_V\cdot\sigma)
\end{equation}
satisfies the traceless condition. Since the first Chern class of
a line bundle must be integral, it follows that $\lambda$ and
$\gamma$ have to obey the following quantization condition
\begin{equation}
\lambda
\gamma+\frac{1}{2}[p_4^{\ast}c_1-c_1(\mathcal{C}^{(4)}_V)]\in
H_{4}(\bar X,\mathbb{Z}).
\end{equation}
With the given cover flux $\gamma$, the net chirality of matter
${\bf 16}$ is calculated by \cite{Donagi:global, Other:global}
\begin{equation}
N_{\bf 16}=(\mathcal{C}^{(4)}_V\cdot\sigma)\cdot\lambda\gamma =
-\lambda\eta\cdot(\eta-4c_1)\label{chirality 16}
\end{equation}
On the other hand, the matter $\bf 10$ corresponds to the
anti-symmetric representation ${\bf 6}$ in $SU(4)_{\bot}$,
associated to the spectral cover $\mathcal{C}^{(6)}_{\wedge^2 V}$.
It turns out that for the $SU(4)$ cover, the net chirality of
matter ${\bf 10}$ is given by \cite{Other:global}
\begin{eqnarray}
N_{\bf 10}= D\cdot\gamma=0.\label{chirality 10}
\end{eqnarray}
It follows from Eqs.~(\ref{chirality 16}) and (\ref{chirality 10})
that one obtain an $SO(10)$ model with
$-\lambda\eta\cdot(\eta-4c_1)$ copies of matter on the ${\bf 16}$
curve and nothing on the $\bf 10$ curve. The flux $\gamma$ does
not have many degrees of freedom to tune and the candidate of $\bf
10$ Higgs is absent. Therefore, in search of realistic models, we
shall consider factorization of the $SU(4)$ cover
$\mathcal{C}_V^{(4)}$ to enrich the configuration, along the line
of the $SU(5)$ cover studied in \cite{Caltech:global02,
Caltech:global03, Blumenhagen:global01, Blumenhagen:global02}. In
the next section, we shall focus on the construction of $(3,1)$
and $(2,2)$ factorizations of the cover $\mathcal{C}_V^{(4)}$.

\section{$SU(4)$ Cover Factorization}

\subsection{$(3,1)$ Factorization}

We consider the $(3,1)$ factorization, $C_V^{(4)} \rightarrow
\mathcal{C}^{(a)}\times\mathcal{C}^{(b)} $ corresponding to the
factorization of Eq. (\ref{homo SU(4) cover}) as follows:
\begin{equation}
\mathcal{C}^{(a)}\times\mathcal{C}^{(b)}:\;\;(a_0 U^3 + a_1U^2V
+a_2 UV^2 + a_3 V^3)(d_0 U +d_1V)=0.
\end{equation}
By comparing with Eq. (\ref{homo SU(4) cover}), one can obtain the
following relations:
\begin{eqnarray}
b_0 = a_0 d_0,~~ b_1 = a_1 d_0 + a_0 d_1,~~ b_2 = a_2 d_0 + a_1
d_1,~~ b_3 = a_3 d_0 + a_2 d_1,~~ b_4 = a_3 d_1.\label{coef 3,1
facorization}
\end{eqnarray}
Let $\xi_1$ be the homological class $[d_1]$ of $d_1$ and write
\begin{equation}
[d_0]= c_1+\xi_1,~~ ~~ [a_k]=\eta -(k+1)c_1-\xi_1,\;\;\;k=0,1,2,3.
\end{equation}
It is easy to see that the homological classes of
$\mathcal{C}^{(a)}$ and $\mathcal{C}^{(b)}$ in $\bar X$ are
\begin{equation}
[\mathcal{C}^{(a)}]= 3\sigma+\pi^{\ast}(\eta
-c_1-\xi_1),\;\;\;[\mathcal{C}^{(b)}]=\sigma+\pi^{\ast}(c_1+\xi_1).\label{31
Classes}
\end{equation}
With the classes given in Eq.~(\ref{31 Classes}), the homological
classes of factorized matter curves $\Sigma_{{\bf 16}^{(a)}}$ and
$\Sigma_{{\bf 16}^{(b)}}$ in $S$ are given by
\begin{equation}
[\Sigma_{{\bf
16}^{(a)}}]=[\mathcal{C}^{(a)}]|_{\sigma}=\eta-4c_1-\xi_1,\;\;\;[\Sigma_{{\bf
16}^{(b)}}]=[\mathcal{C}^{(b)}]|_{\sigma}=\xi_1.
\end{equation}
To obtain the factorized ${\bf 10}$ curves, we follow the method
proposed in \cite{Donagi:2004ia, Caltech:global02,
Caltech:global03, Blumenhagen:global01} to calculate the
intersection $\mathcal{C}^{(4)}_V\cap \tau \mathcal{C}^{(4)}_V$,
where $\tau$ is the $\mathbb{Z}_2$ involution $\tau: V\rightarrow
-V$ acting on the spectral cover. Since the calculation is
straightforward, we omit the detailed calculation here and only
summarize the results\footnote{To simplify notations, we denote
$\mathcal{C}^{(k)}\cap\tau\mathcal{C}^{(l)}$ by
$\mathcal{C}^{(k)(l)}$. Notice that
$[\mathcal{C}^{(k)(l)}]=[\mathcal{C}^{(l)(k)}]$.} in Table
\ref{3-1 10curve}.

\begin{table}[h]
\begin{center}
\renewcommand{\arraystretch}{.75}
\begin{tabular}{|c|c|c|c|} \hline
& $[\mathcal{C}^{(b)(b)}]$ & $2[\mathcal{C}^{(a)(b)}]$ & $[\mathcal{C}^{(a)(a)}]$ \\
\hline

{\bf 16} & $\sigma\cdot\pi^{\ast}\xi_1$ & - &
$\sigma\cdot\pi^{\ast} (\eta-4c_1-\xi_1)$ \\ \hline

\multirow{2}{*}{\bf 10} & \multirow{2}{*}{$\pi^{\ast}
\xi_1\cdot\pi^{\ast} (c_1+\xi_1)$} &
$2[\sigma+\pi^{\ast}(c_1+\xi_1)]$ &
$[2\sigma+\pi^{\ast}(\eta-2c_1 -\xi_1)]$\\
& & $\cdot~\pi^{\ast}(\eta-3c_1-\xi_1)+2\sigma\cdot
\pi^{\ast}\xi_1$ & $\cdot~\pi^{\ast}(\eta
-3c_1-\xi_1)+2(\sigma+\pi^{\ast}c_1)\cdot \pi^{\ast} \xi_1$ \\
\hline

\multirow{2}{*}{$\infty$} & \multirow{2}{*}{$\sigma_{\infty}
\cdot\pi^{\ast} (c_1+\xi_1)$} & \multirow{2}{*}{4$\sigma_{\infty}
\cdot \pi^{\ast} (c_1+\xi_1)$} & $\sigma_{\infty} \cdot
\pi^{\ast}(\eta-c_1 -\xi_1)$ \\
& & & $+2\sigma_{\infty} \cdot \pi^{\ast}(\eta -2c_1-2\xi_1)$
\\ \hline
\end{tabular}
\caption{The homological classes of the matter curves in the
$(3,1)$ factorization.} \label{3-1 10curve}
\end{center}
\end{table}

It follows from Table \ref{3-1 10curve} that the relevant
classes %\footnote{$[\mathcal{C}^{(b)(b)}]$ is a point in $S$.}
in $\bar X$ for $\bf 10$ curves are
\begin{eqnarray}
&&[\mathcal{C}^{(a)(a)}]=[2\sigma+\pi^{\ast}(\eta-2c_1
-\xi_1)]\cdot~\pi^{\ast}(\eta
-3c_1-\xi_1)+2(\sigma+\pi^{\ast}c_1)\cdot \pi^{\ast} \xi_1,~~~~ \\
&&[\mathcal{C}^{(a)(b)}]=[\sigma+\pi^{\ast}(c_1+\xi_1)]\cdot~\pi^{\ast}(\eta-3c_1-\xi_1)+\sigma\cdot
\pi^{\ast}\xi_1,
\end{eqnarray}
which give rise to the $\bf 10$ curves
\begin{equation}
[\Sigma_{{\bf 10}^{(a)(a)}}]=\eta-3c_1,\;\;\;[\Sigma_{{\bf
10}^{(a)(b)}}]=\eta-3c_1\label{31 matter curve 10 in S},
\end{equation}
respectively%\footnote{Alternatively, since the $\bf 10$ curves correspond to $b_3=0$, it follows from Eq. (\ref{coef 3,1
%facorization}) that the corresponding decomposed covers are:
%$\mathcal{C}_{\wedge^2 V}^{(a)(a)}: a_0 d_0 s^3 - 2 a_0 d_1 s^2 +
%(a_2 d_0-a_1d_1)s -(a_2 d_1+a_3 d_0)=0$ and
%$\mathcal{C}_{\wedge^2V}^{(a)(b)}: a_0 d_0 s^3 +(a_2 d_0-a_1d_1)s
%+(a_2 d_1 + a_3 d_0)=0$, which also leads to Eq. (\ref{31 matter
%curve 10 in S}). Note that $[\mathcal{C}_{\wedge^2
%V}^{(a)(a)}]\cap\sigma\neq [\mathcal{C}^{(a)(a)}]\cap\sigma$ and
%$[\mathcal{C}_{\wedge^2 V}^{(a)(b)}]\cap\sigma\neq
%[\mathcal{C}^{(a)(b)}]\cap\sigma$. However,
%$[\mathcal{C}_{\wedge^2
%V}^{(a)(a)}]|_{\sigma}=[\mathcal{C}^{(a)(a)}]|_{\sigma}$ and
%$[\mathcal{C}_{\wedge^2 V}^{(a)(b)}]|_{\sigma}=
%[\mathcal{C}^{(a)(b)}]|_{\sigma}$.}
.

\subsection{$(2,2)$ Factorization}

In the $(2,2)$ factorization, the cover is split as
$\mathcal{C}^{(4)}_V \rightarrow \mathcal{C}^{(d_1)}\times
\mathcal{C}^{(d_2)}$. More precisely, the cover defined in
Eq.~(\ref{homo SU(4) cover}) is factorized into the following
form:
\begin{equation}
\mathcal{C}^{(d_1)}\times \mathcal{C}^{(d_2)}:\;\;\;(e_0U^2 +
e_1UV +e_2V^2)(f_0U^2 + f_1UV + f_2V^2) =0.
\end{equation}
By comparing the coefficients with Eq. (\ref{homo SU(4) cover}),
one obtains
\begin{equation}
b_0=e_0f_0,~~ b_1=e_0f_1+e_1f_0,~~ b_2=e_0f_2+e_1f_1+e_2f_0,~~ b_3
= e_1f_2+ e_2f_1,~~ b_4=e_2f_2.
\end{equation}
Let $\xi_2$ be the homological class of $f_2$ and then the
homological classes of other sections can be written as
\begin{eqnarray}
[f_1]= c_1+\xi_2, ~~ [f_0]= 2c_1+\xi_2, ~~[e_m]= \eta-(m+2)c_1 -
\xi_2,\;\;\;m=0,1,2.
\end{eqnarray}
In this case, the homological classes of $\mathcal{C}^{(d_1)}$ and
$\mathcal{C}^{(d_2)}$ are given by
\begin{equation}
[\mathcal{C}^{(d_1)}]= 2\sigma+\pi^{\ast}(\eta
-2c_1-\xi_2),\;\;\;[\mathcal{C}^{(d_2)}]=2\sigma+\pi^{\ast}(2c_1+\xi_2).
\end{equation}
The homological classes of the corresponding matter curves
$\Sigma_{{\bf 16}^{(d_1)}}$ and $\Sigma_{{\bf 16}^{(d_2)}}$ are
then computed as
\begin{equation}
[\Sigma_{{\bf
16}^{(d_1)}}]=[\mathcal{C}^{(d_1)}]|_{\sigma}=\eta-4c_1-\xi_2,\;\;\;[\Sigma_{{\bf
16}^{(d_2)}}]=[\mathcal{C}^{(d_2)}]|_{\sigma}=\xi_2,
\end{equation}
respectively. To calculate the homological classes of the
factorized ${\bf 10}$ curves, we again follow the method proposed
in \cite{Donagi:2004ia, Caltech:global02, Caltech:global03,
Blumenhagen:global01} to calculate the intersection
$\mathcal{C}^{(4)}_V\cap \tau \mathcal{C}^{(4)}_V$. We omit the
detailed calculation here and only summarize the
results %\footnote{In this case, we may impose an ansatz $e_0=\alpha f_0$ and $e_1=-\alpha f_1$.}
in Table \ref{2-2 10curve}.
\begin{table}[h]
\begin{center}
\renewcommand{\arraystretch}{.75}
\begin{tabular}{|c|c|c|c|} \hline
& $[\mathcal{C}^{(d_2)(d_2)}]$ & $2[\mathcal{C}^{(d_1)(d_2)}]$ & $[\mathcal{C}^{(d_1)(d_1)}]$ \\
\hline

{\bf 16} & $\sigma\cdot\pi^{\ast}\xi_2$ & - &
$\sigma\cdot\pi^{\ast} (\eta-4c_1-\xi_2)$ \\ \hline

\multirow{2}{*}{\bf 10} & $(2\sigma +\pi^{\ast}( 2c_1+\xi_2))$ &
$2(2\sigma+\pi^{\ast}(2c_1+\xi_2))$ &
$\pi^{\ast}(\eta-3c_1-\xi_2)\cdot \pi^{\ast}(\eta-4c_1 -\xi_2)$\\
&$\cdot~\pi^{\ast} (c_1+\xi_2)$ &
$\cdot~\pi^{\ast}(\eta-4c_1-\xi_2)$ & $+2(\sigma+\pi^{\ast}c_1
)\cdot \pi^{\ast}(c_1+\xi_2)$ \\ \hline

\multirow{2}{*}{$\infty$} & \multirow{2}{*}{$\sigma_{\infty}
\cdot\pi^{\ast} (2c_1+\xi_2)$} & \multirow{2}{*}{4$\sigma_{\infty}
\cdot \pi^{\ast} (2c_1+\xi_2)$} & $\sigma_{\infty} \cdot
\pi^{\ast}(\eta-2c_1 -\xi_2)$ \\
& & & $+2\sigma_{\infty} \cdot \pi^{\ast}(\eta -4c_1-2\xi_2)$
\\ \hline
\end{tabular}
\caption{The homological classes of the matter curves in the
$(2,2)$ factorization.} \label{2-2 10curve}
\end{center}
\end{table}

It follows from Table \ref{2-2 10curve} that the classes in $\bar
X$ for the factorized $\bf 10$ curves are as follows:
\begin{eqnarray}
&&[\mathcal{C}^{(d_1)(d_1)}]=2(\sigma+\pi^{\ast}c_1 )\cdot
\pi^{\ast}(c_1+\xi_2)+\pi^{\ast}(\eta-3c_1-\xi_2)\cdot
\pi^{\ast}(\eta-4c_1 -\xi_2),~~~~~~~~ \\
&&[\mathcal{C}^{(d_1)(d_2)}]=(2\sigma+\pi^{\ast}(2c_1+\xi_2))\cdot~\pi^{\ast}(\eta-4c_1-\xi_2),
\\
&&[\mathcal{C}^{(d_2)(d_2)}]=(2\sigma+\pi^{\ast}(2c_1+\xi_2))\cdot~\pi^{\ast}(c_1+\xi_2).
\end{eqnarray}
With the classes $[\mathcal{C}^{(d_1)(d_1)}]$,
$[\mathcal{C}^{(d_1)(d_2)}]$, and $[\mathcal{C}^{(d_2)(d_2)}]$,
one can calculate the classes of the corresponding $\bf 10$ curves
in $S$
as follows%\footnote{Alternatively, since the $\bf 10$ curves
%correspond to $b_3=0$, it follows from Eq. (\ref{coef 3,1
%facorization}) that $b_3= e_2 f_1 + e_1 f_2 = f_1(e_2 -\alpha
%f_2)$, so we obtain two kinds of $\bf 10$ curve,
%$[b_3]_1=\pi^{\ast} (c_1 +\xi)$ and $[b_3]_2 = \pi^{\ast}
%(\eta-4c_1-\xi)$. The corresponding decomposed covers are:
%$\mathcal{C}_{\wedge^2 V}^{(d_1)(d_1)}: e_0 s + e_1=0$,
%$\mathcal{C}_{\wedge^2 V}^{(d_2)(d_2)}: f_0 s + f_1=0$, and
%$\mathcal{C}_{\wedge^2 V}^{(d_1)(d_2)}: e_0^2 f_0^2 s^4 +e_0 f_0
%(2e_2 f_0 + e_1 f_1 + 2e_0 f_2)s^2 +(e_2 f_0-e_0 f_2)^2=0$. By
%using the ansatz $e_0=\alpha f_0$ and $e_1=-\alpha f_1$, we have
%$\mathcal{C}_{\wedge^2 V}^{(d_1)(d_1)}: \alpha(f_0 s - f_1)=0$, $
%\mathcal{C}_{\wedge^2 V}^{(d_2)(d_2)}: f_0 s + f_1=0$, and
%$\mathcal{C}_{\wedge^2 V}^{(d_1)(d_2)}: f_0^2[\alpha^2 f_0^2 s^4 +
%\alpha(2e_2 f_0 -\alpha f_1^2 +2\alpha f_0 f_2)s +(e_2 -\alpha f_2)^2]=0$, which also gives rise
%to Eq. (\ref{22 matter curve 10 in S}).}
:
\begin{equation}
[\Sigma_{{\bf 10}^{(d_1)(d_1)}}]=c_1+\xi_2,\;\;\;[\Sigma_{{\bf
10}^{(d_1)(d_2)}}]=2\eta-8c_1-2\xi_2,\;\;\; [\Sigma_{{\bf
10}^{(d_2)(d_2)}}]=c_1+\xi_2 . \label{22 matter curve 10 in S}
\end{equation}

\section{Spectral Cover Fluxes}

Let us consider the case of the cover factorization
$\mathcal{C}_{V}^{(n)}\ra\mathcal{C}^{(l)}\times\mathcal{C}^{(m)}$.
To obtain well-defined cover fluxes and maintain supersymmetry, we
impose the following constraints \cite{Caltech:global03}:
\begin{eqnarray}
&&c_1(p_{l\ast}\mathcal{L}^{(l)}) +
c_1(p_{m\ast}\mathcal{L}^{(m)})
=0,\label{cover constraint 01}\\
&&c_1(\mathcal{L}^{(k)})\in
H_2(\mathcal{C}^{(k)},\mathbb{Z}),\;\;k=l,m,\label{cover
constraint
02}\\
&&[c_1(p_{l\ast}\mathcal{L}^{(l)}) -
c_1(p_{m\ast}\mathcal{L}^{(m)})]\cdot_S [\omega]=0,\label{cover
constraint 03}
\end{eqnarray}
where $p_k$ denotes the projection map from the cover
$\mathcal{C}^{(k)}$ to $S$, $p_k:~\mathcal{C}^{(k)}\rightarrow S$,
$\mathcal{L}^{(k)}$ is a line bundle over $\mathcal{C}^{(k)}$ and
$[\omega]$ is an ample divisor dual to a K\"ahler form of $S$. The
first constraint Eq.~(\ref{cover constraint 01}) is the traceless
condition for the induced Higgs bundle\footnote{One may think of
Eq. (\ref{cover constraint 01}) as the traceless condition of an
$SU(4)$ bundle $V_4$ over $S$ split into $V_3\oplus L$ with
$V_3=p_{a\ast}\mathcal{L}^{(a)}$ and
$L=p_{b\ast}\mathcal{L}^{(b)}$. Then the traceless condition of
$V_4$ can be expressed by $c_1(V_4)
=c_1(p_{a\ast}\mathcal{L}^{(a)}) +
c_1(p_{b\ast}\mathcal{L}^{(b)})=0$.}.  The second constraint Eq.
(\ref{cover constraint 02}) requires that the first Chern class of
a well-defined line bundle $\mathcal{L}^{(k)}$ over
$\mathcal{C}^{(k)}$ must be integral. The third constraint states
that the 2-cycle $c_1(p_{l\ast}\mathcal{L}^{(l)}) -
c_1(p_{m\ast}\mathcal{L}^{(m)})$ in $S$ has to be supersymmetic.
Note that Eq. (\ref{cover constraint 01}) can be expressed as
\begin{equation}
p_{l\ast}c_1(\mathcal{L}^{(l)})-\frac{1}{2}p_{l\ast}r^{(l)} +
p_{m\ast}c_1(\mathcal{L}^{(m)})-\frac{1}{2}p_{m\ast}r^{(m)}=0,
\end{equation}
where $r^{(l)}$ and $r^{(m)}$ are the ramification divisors for
the maps $p_l$ and $p_m$, respectively. Recall that the
ramification divisors $r^{(k)}$ are defined by
\begin{equation}
r^{(k)}=p_k^{\ast} c_1 -c_1(\mathcal{C}^{(k)}),\;\;k=l,m.\label{r
divisor}
\end{equation}
The term $c_1(\mathcal{C}^{(k)})$ in Eq. (\ref{r divisor}) can be
calculated by the adjuction formula \cite{Griffith:01,
Hartshone:01},
\begin{equation}
c_1(\mathcal{C}^{(k)})=(c_1(\bar X)-[\mathcal{C}^{(k)}])\cdot
[\mathcal{C}^{(k)}].
\end{equation}
It is convenient to define cover fluxes $\gamma^{(k)}$ as
\begin{equation}
c_1(\mathcal{L}^{(k)})=\gamma^{(k)}+\frac{1}{2}
r^{(k)},\;\;k=l,m.\label{cover flux def}
\end{equation}
With Eq. (\ref{cover flux def}), the traceless condition Eq.
(\ref{cover constraint 01}) can be expressed as
$p_{l\ast}\gamma^{(l)}+p_{m\ast}\gamma^{(m)}=0$. By using Eq.
(\ref{r divisor}) and Eq. (\ref{cover flux def}), we can recast
the quantization condition Eq. (\ref{cover constraint 02}) by
$\gamma^{(k)}+\frac{1}{2}[p^{\ast}_{k}c_1-c_1(\mathcal{C}^{(k)})]\in
H_2(\mathcal{C}^{(k)},\mathbb{Z}),\;\;k=l,m$. Finally, the
supersymmetry condition Eq. (\ref{cover constraint 03}) is reduced
to $p_{k\ast}\gamma^{(k)}\cdot_S[\omega]=0$. We summarize the
constraints as follows:
\begin{eqnarray}
&& p_{l\ast}\gamma^{(l)}+p_{m\ast}\gamma^{(m)}=0,\label{simplified
cover constraint 01}
\\
&&\gamma^{(k)}+\frac{1}{2}[p^{\ast}_{k}c_1-c_1(\mathcal{C}^{(k)})]\in
H_2(\mathcal{C}^{(k)},\mathbb{Z}),\;\;k=l,m,\label{simplified
cover constraint 02}
\\
&&
p_{k\ast}\gamma^{(k)}\cdot_S[\omega]=0,\;\;k=l,m.\label{simplified
cover constraint 03}
\end{eqnarray}
In the next section, we shall explicitly construct the cover
fluxes $\gamma^{(k)}$ satisfying Eq. (\ref{simplified cover
constraint 01}), (\ref{simplified cover constraint 02}), and
(\ref{simplified cover constraint 03}) for the $(3,1)$ and $(2,2)$
factorizations. We also calculate the restrictions of the fluxes
to each matter curve.

\subsection{(3,1) Factorization}

In the $(3,1)$ factorization, the ramification divisors for the
spectral covers $\mathcal{C}^{(a)}$ and $\mathcal{C}^{(b)}$ are
given by
\begin{eqnarray}
&&r^{(a)}=[\mathcal{C}^{(a)}]\cdot[\sigma+\pi^{\ast}(\eta-2c_1-\xi_1)], \\
&&r^{(b)}=[\mathcal{C}^{(b)}]\cdot (-\sigma+\pi^{\ast}\xi_1),
\end{eqnarray}
respectively. We define traceless fluxes $\gamma^{(a)}_0$ and
$\gamma^{(b)}_0$ by
\begin{eqnarray}
&&\gamma^{(a)}_0=(3-p_a^{\ast} p_{a\ast}) \gamma^{(a)}
=[\mathcal{C}^{(a)}]\cdot [ 3
\sigma-\pi^{\ast}(\eta-4c_1-\xi_1)],\\
&&\gamma^{(b)}_0=(1-p_b^{\ast} p_{b\ast})\gamma^{(b)}
=[\mathcal{C}^{(b)}]\cdot \left(
\sigma-\pi^{\ast}\xi_1\right),\label{Flux_1_3,1}
\end{eqnarray}
where $\gamma^{(a)}$ and $\gamma^{(b)}$ are non-traceless fluxes
and defined as
\begin{equation}
\gamma^{(a)}=[\mathcal{C}^{(a)}]\cdot \sigma,~~
\gamma^{(b)}=[\mathcal{C}^{(b)}]\cdot \sigma.
\end{equation}
Then we can  calculate the restriction of fluxes $\gamma^{(a)}_0$
and $\gamma^{(b)}_0$ to each matter curve. We omit the calculation
here and only summarize the results in the following table.
\begin{equation}
\begin{tabular}{lcc} \hline
& $\gamma^{(b)}_0$ & $\gamma^{(a)}_0$ \\\hline

${\bf 16}^{(b)}$ & $-\xi_1\cdot_S(c_1+\xi_1)$ & 0 \\ \hline

${\bf 16}^{(a)}$ & 0 & $-(\eta-c_1-\xi_1)\cdot_S(\eta-4c_1-\xi_1)$
\\\hline

${\bf 10}^{(a)(b)}$ & $-\xi_1\cdot_S(c_1+\xi_1)$ &
$-(\eta-3c_1-3\xi_1)\cdot_S(\eta-4c_1-\xi_1)$ \\\hline

${\bf 10}^{(a)(a)}$ & 0 &
$(\eta-3c_1-3\xi_1)\cdot_S(\eta-4c_1-\xi_1)$
\\\hline
\end{tabular}
\end{equation}
Due to the factorization, one also can define additional fluxes
$\delta^{(a)}$ and $\delta^{(b)}$ by
\begin{eqnarray}
&&\delta^{(a)}=(1-p_b^{\ast} p_{a\ast}) \gamma^{(a)}
=[\mathcal{C}^{(a)}]\cdot \sigma-[\mathcal{C}^{(b)}]\cdot
\pi^{\ast}(\eta-4c_1-\xi_1) \nonumber
\\&&\delta^{(b)}=(3-p_a^{\ast} p_{b\ast}) \gamma^{(b)}
=[\mathcal{C}^{(b)}]\cdot 3\sigma-
[\mathcal{C}^{(a)}]\cdot\pi^{\ast}\xi_1. \label{Flux_2_3,1}
\end{eqnarray}
Another flux one can include is \cite{Caltech:global03}
\begin{equation}
\tilde{\rho}=(3p_{b}^{\ast}-p_{a}^{\ast})\rho,\label{Flux_3_3,1}
\end{equation}
for any $\rho\in H_2(S,\mathbb{R})$.  We summarize the restriction
of fluxes $\delta^{(a)}$, $\delta^{(b)}$ and $\tilde{\rho}$ to
each matter curve in the following table.
\begin{equation}
\begin{tabular}{lccc} \hline
& $\delta^{(b)}$ & $\delta^{(a)}$ & $\tilde{\rho}$ \\\hline

${\bf 16}^{(b)}$ & $-3c_1\cdot_S\xi_1$ &
$-\xi_1\cdot_S(\eta-4c_1-\xi_1)$ & $3\rho\cdot_S \xi_1$ \\ \hline

${\bf 16}^{(a)}$ & $-\xi_1\cdot_S (\eta-4c_1-\xi_1)$ &
$-c_1\cdot_S(\eta-4c_1-\xi_1)$ & $-\rho\cdot_S (\eta-4c_1-\xi_1)$
\\\hline

${\bf 10}^{(a)(b)}$ & $\xi_1\cdot_S(2\eta-9c_1-3\xi_1)$ &
$-(\eta-3c_1-\xi_1)\cdot_S(\eta-4c_1-\xi_1)$&
$2\rho\cdot_S(\eta-3c_1)$
\\\hline

${\bf 10}^{(a)(a)}$ & $-2\xi_1\cdot_S (\eta-3c_1)$ &
$(\eta-3c_1-\xi_1)\cdot_S(\eta-4c_1-\xi_1)$ & $-2\rho\cdot_S
(\eta-3c_1)$
\\\hline
\end{tabular}
\end{equation}

With Eqs.~(\ref{Flux_1_3,1}), (\ref{Flux_2_3,1}), and
(\ref{Flux_3_3,1}), we define the universal cover flux $\Gamma$ to
be \cite{Caltech:global03}
\begin{equation}
\Gamma=k_a \gamma^{(a)}_0+k_b\gamma^{(b)}_0 +m_a \delta^{(a)} +m_b
\delta^{(b)} + \tilde{\rho}\equiv\Gamma^{(a)}+\Gamma^{(b)},
\end{equation}
where $\Gamma^{(a)}$ and $\Gamma^{(b)}$ are given by
\begin{eqnarray}
&&\Gamma^{(a)}=[\mathcal{C}^{(a)}]\cdot \left[(3k_a+m_a)\sigma
-\pi^{\ast}(k_a (\eta-4c_1-\xi_1) +m_b\xi_1 + \rho)\right],
\\&&\Gamma^{(b)}=[\mathcal{C}^{(b)}]\cdot \left[(k_b+3m_b)\sigma
-\pi^{\ast}(k_b\xi_1+m_a(\eta-4c_1-\xi_1)-3\rho) \right].
\end{eqnarray}
Note that
\begin{eqnarray}
&&p_{a\ast}\Gamma^{(a)}= -3m_b\xi_1+m_a(\eta-4c_1-\xi_1)-3\rho, \\
&&p_{b\ast} \Gamma^{(b)}= 3m_b\xi_1 -m_a(\eta-4c_1-\xi_1)+3\rho.
\end{eqnarray}
Clearly, $\Gamma^{(a)}$ and $\Gamma^{(b)}$ obey the traceless
condition $p_{a\ast}\Gamma^{(a)}+p_{b\ast}\Gamma^{(b)}=0$.
Besides, the quantization condition in this case becomes
\begin{equation}
(3k_a+m_a+\frac{1}{2})\sigma -\pi^{\ast}[k_a(\eta-4c_1-\xi_1)+
m_b\xi_1 + \rho-\frac{1}{2}(\eta-2c_1-\xi_1)] \in H_4({\bar
X},\mathbb{Z}),
\end{equation}
\begin{equation}
(k_b+3m_b-\frac{1}{2})\sigma -\pi^{\ast}[k_b\xi_1
+m_a(\eta-4c_1-\xi_1)-3\rho -\frac{1}{2}\xi_1] \in H_4({\bar
X},\mathbb{Z}).
\end{equation}
The supersymmetry condition is given by
\begin{equation}
[3m_b\xi_1-m_a(\eta-4c_1-\xi_1)+3\rho]\cdot_S[\omega]=0.\label{BPS
3,1}
\end{equation}

\subsection{(2,2) Factorization}

We can calculate the ramification divisors $r^{(d_1)}$ and
$r^{(d_2)}$ for the $(2,2)$ factorization and obtain
\begin{eqnarray}
&&r^{(d_1)}= [\mathcal{C}^{(d_1)}]\cdot\pi^{\ast}(\eta-3c_1-\xi_2),  \\
&&r^{(d_2)}= [\mathcal{C}^{(d_2)}]\cdot\pi^{\ast}(c_1+\xi_2).
\end{eqnarray}
We then define traceless cover fluxes $\gamma^{(d_1)}_0$ and
$\gamma^{(d_2)}_0$ by
\begin{eqnarray}
&&\gamma^{(d_1)}_0=(2-p_{d_1}^{\ast} p_{d_1\ast})\gamma^{(d_1)}
=[\mathcal{C}^{(d_1)}]\cdot \left[ 2
\sigma-\pi^{\ast}(\eta-4c_1-\xi_2)\right], \\
&&\gamma^{(d_2)}_0=(2-p_{d_2}^{\ast} p_{d_2\ast})\gamma^{(d_2)}
=[\mathcal{C}^{(d_2)}]\cdot \left( 2
\sigma-\pi^{\ast}\xi_2\right),
\end{eqnarray}
where $\gamma^{(d_1)}$ and $\gamma^{(d_21)}$ are non-traceless
fluxes and given by
\begin{equation}
\gamma^{(d_1)}=[\mathcal{C}^{(d_1)}]\cdot \sigma,~~
\gamma^{(d_2)}=[\mathcal{C}^{(d_2)}]\cdot \sigma.
\end{equation}
We summarize the restriction of the fluxes to each factorized
curve in the following table.
\begin{equation}
\begin{tabular}{ccc} \hline
& $\gamma^{(d_2)}_0$ & $\gamma^{(d_1)}_0$ \\\hline

${\bf 16}^{(d_2)}$ & $-\xi_2\cdot_S(2c_1+\xi_2)$ & 0 \\ \hline

${\bf 16}^{(d_1)}$ & 0 &
$-(\eta-2c_1-\xi_2)\cdot_S(\eta-4c_1-\xi_2)$ \\\hline

${\bf 10}^{(d_2)(d_2)}$ & 0 & 0 \\ \hline

${\bf 10}^{(d_1)(d_2)}$ & 0 &
$-2(\eta-4c_1-2\xi_2)\cdot_S(\eta-4c_1-\xi_2)$
\\\hline

${\bf 10}^{(d_1)(d_1)}$ & 0 &
$2(\eta-4c_1-2\xi_2)\cdot_S(\eta-4c_1-\xi_2)$
\\\hline
\end{tabular}
\end{equation}
Due to the factorization, one also can define following fluxes
\cite{Caltech:global03}
\begin{eqnarray}
&&\delta^{(d_1)}=(2-p_{d_2}^{\ast} p_{d_1\ast}) \gamma^{(d_1)}
=[\mathcal{C}^{(d_1)}]\cdot
2\sigma- [\mathcal{C}^{(d_2)}]\cdot\pi^{\ast}(\eta-4c_1-\xi_2), \nonumber \\
&&\delta^{(d_2)}=(2-p_{d_1}^{\ast} p_{d_2\ast}) \gamma^{(d_2)}
=[\mathcal{C}^{(d_2)}]\cdot 2\sigma-[\mathcal{C}^{(d_1)}]\cdot
\pi^{\ast}\xi_2,
\end{eqnarray}
and
\begin{equation}
\widehat{\rho}=(p_{d_2}^{\ast}-p_{d_1}^{\ast})\rho,
\end{equation}
for any $\rho\in H_2(S,\mathbb{R})$. We summarize the restriction
of the fluxes $\delta^{(d_1)}$, $\delta^{(d_2)}$, and
$\widehat{{\rho}}$ to each factorized curve as follows:
\begin{equation}
\begin{tabular}{lccc} \hline
& $\delta^{(d_2)}$ & $\delta^{(d_1)}$ & $\widehat{{\rho}}$
\\\hline

${{\bf 16}^{(d_2)}}$ & $-2c_1\cdot_S\xi_2$ & $-\xi_2\cdot_S
(\eta-4c_1-\xi_2)$ & $\rho\cdot_S \xi_2$ \\ \hline

${{\bf 16}^{(d_1)}}$ & $-\xi_2\cdot_S (\eta-4c_1-\xi_2)$ &
$-2c_1\cdot_S(\eta-4c_1-\xi_2)$ & $-\rho\cdot_S (\eta-4c_1-\xi_2)$
\\\hline

${{\bf 10}^{(d_2)(d_2)}}$ & $2\xi_2\cdot_S (c_1+\xi_2)$ &
$-2(c_1+\xi_2)\cdot_S(\eta-4c_1-\xi_2)$ & $2\rho\cdot_S(c_1+\xi_2)$\\
\hline

${{\bf 10}^{(d_1)(d_2)}}$ & 0 &
$-2(\eta-4c_1-2\xi_2)\cdot_S(\eta-4c_1-\xi_2)$&0
\\\hline

${{\bf 10}^{(d_1)(d_1)}}$ & $-2\xi_2\cdot_S(c_1+\xi_2)$ &
$2(\eta-3c_1-\xi_2)\cdot_S(\eta-4c_1-\xi_2)$ & $-2\rho\cdot_S
(c_1+\xi_2)$
\\\hline
\end{tabular}
\end{equation}

In this case the universal cover flux is defined by
\begin{equation}
\Gamma=k_{d_1} \gamma^{(d_1)}_0+k_{d_2}\gamma^{(d_2)}_0 +m_{d_1}
\delta^{(d_1)} +m_{d_2} \delta^{(d_2)} +
\widehat{{\rho}}=\Gamma^{(d_1)}+\Gamma^{(d_2)},
\end{equation}
where
\begin{eqnarray}
&&\Gamma^{(d_1)}=[\mathcal{C}^{(d_1)}]\cdot
\left\{2(k_{d_1}+m_{d_1})\sigma -\pi^{\ast}[k_{d_1}
(\eta-4c_1-\xi_2)+m_{d_2}\xi_2+ \rho]\right\},\nonumber \\
&&\Gamma^{(d_2)}=[\mathcal{C}^{(d_2)}]\cdot
\left\{2(k_{d_2}+m_{d_2})\sigma
-\pi^{\ast}[k_{d_2}\xi_2+m_{d_1}(\eta-4c_1-\xi_2)-\rho] \right\} .
\end{eqnarray}
Note that
\begin{eqnarray}
&&p_{d_1\ast} \Gamma^{(d_1)}
=-2m_{d_2}\xi_2+2m_{d_1}(\eta-4c_1-\xi_2)-2\rho,\\
&&p_{d_2\ast} \Gamma^{(d_2)} =
2m_{d_2}\xi_2-2m_{d_1}(\eta-4c_1-\xi_2)+2\rho.
\end{eqnarray}
It is easy to see that $\Gamma^{(d_1)}$ and $\Gamma^{(d_2)}$
satisfy the traceless condition $p_{d_1\ast}
\Gamma^{(d_1)}+p_{d_2\ast} \Gamma^{(d_2)}=0$. In addition, the
quantization condition in this case becomes
\begin{equation}
2(k_{d_1}+m_{d_1})\sigma -\pi^{\ast}[k_{d_1}(\eta-4c_1-\xi_2)+
m_{d_2}\xi_2 + \rho-\frac{1}{2}(\eta-3c_1-\xi_2)] \in H_4({\bar
X},\mathbb{Z}),
\end{equation}
\begin{equation}
2(k_{d_2}+m_{d_2})\sigma
-\pi^{\ast}[k_{d_2}\xi_2+m_{d_1}(\eta-4c_1-\xi_2)-\rho
-\frac{1}{2}(c_1+\xi_2)]\in H_4({\bar X},\mathbb{Z}).
\end{equation}
The supersymmetry condition is then given by
\begin{equation}
[2m_{d_2}\xi_2-2m_{d_1}(\eta-4c_1-\xi_2)+2\rho]\cdot_S[\omega]=0.\label{BPS
2,2}
\end{equation}

\section{$D3$-brane Tadpole Cancellation}

The cancellation of tadpoles is crucial for consistent
compactifications. In general, there are induced tadpoles from
7-brane, 5-brane, and 3-brane charges in F-theory. It is well
known that 7-brane tadpole cancellation in F-theory is
automatically satisfied since $X_4$ is a Calabi-Yau manifold. In
spectral cover models, the cancellation of the $D5$-brane tadpole
follows from the topological condition that the overall first
Chern class of the Higgs bundle vanishes. Therefore, the
non-trivial tadpole cancellation needed to be satisfied is the
$D3$-brane tadpole. The $D3$-brane tadpole can be calculated by
the Euler characteristic $\chi(X_4)$. The cancellation condition
is of the form \cite{Sethi:1996es}
\begin{equation}
N_{D3}=\frac{{\chi}(X_4)}{24}-\frac{1}{2}\int_{X_4}G\wedge
G,\label{Tadpole cancellation}
\end{equation}
where $N_{D3}$ is the number of $D3$-branes and $G$ is the
four-form flux on $X_4$. For a non-singular elliptically fibered
Calabi-Yau manifold, it was shown in \cite{Sethi:1996es} that the
Euler characteristic $\chi(X_4)$ can be expressed as
\begin{equation}
\chi(X_4)=12\int_{B_3}c_1(B_3)[c_2(B_3)+30c_1(B_3)^2],\label{Euler
number smooth}
\end{equation}
where $c_k(B_3)$ are the Chern classes of $B_3$. It follows from
Eq. (\ref{Euler number smooth}) that $\chi(X_4)/24$ is at least
half-integral\footnote{For a generic Calabi-Yau manifold, it was
shown in \cite{Sethi:1996es} that $\chi(X_4)/6\in \mathbb{Z}$,
which implies that $\chi(X_4)/24$ takes value in
$\mathbb{\mathbb{Z}}_4$.}. When $X_4$ admits non-abelian
singularities, the Euler characteristic of $X_4$ is replaced by
the refined Euler characteristic, the Euler characteristic of the
smooth fourfold obtained from a suitable resolution of $X_4$. On
the other hand, $G$-flux encodes the two-form gauge fluxes on
7-branes. It was shown in \cite{Curio:1998bva} that
\begin{equation}
\frac{1}{2}\int_{X_4} G\wedge
G=-\frac{1}{2}\Gamma^{2},\label{self-intersection Gamma}
\end{equation}
where $\Gamma$ is the universal cover flux defined in section 4
and $\Gamma^2$ is the self-intersection number of $\Gamma$ inside
the spectral cover\footnote{Eq.~(\ref{self-intersection Gamma})
originates from the spectral cover construction in heterotic
string compactifications \cite{Curio:1998bva}. This equation holds
for F-theory compactified on elliptically fibered fourfolds
possessing a heterotic dual by heterotic/F-theory duality.
However, since $X_4$ is not a global fibration over $S$, we assume
that Eq.~(\ref{self-intersection Gamma}) is valid for F-theory
models without heterotic dual, and the fluxes can correctly
described by spectral covers. }. It is a challenge to find
compactifications with non-vanishing $G$-flux and non-negative
$N_{D_3}$ to satisfy the tadpole cancellation condition
Eq.~(\ref{Tadpole cancellation}). In the next two subsections, we
shall derive the formulae of refined Euler characteristic
$\chi(X_4)$ and the self-intersection of universal cover fluxes
$\Gamma^2$ for $(3,1)$ and $(2,2)$ factorizations.

\subsection{Geometric Contribution}

In the presence of non-abelian singularities, $X_4$ becomes
singular and the Euler characteristic $\chi(X_4)$ is modified by
resolving the singularities. To be more concrete, let us consider
$X_4$ with an elliptic fibration which degenerates over $S$ to a
non-abelian singularity corresponding to gauge group $H$ and
define $G$ to be the complement of $H$ in $E_8$. The Euler
characteristic is modified to
\begin{equation}
{\chi}(X_4)=\chi^{\ast}(X_4)+\chi_G-\chi_{E_8},\label{refined chi}
\end{equation}
where $\chi^{\ast}(X_4)$ is the Euler characteristic for a smooth
fibration over $B_3$ given by Eq. (\ref{Euler number smooth}). The
characteristic $\chi_{E_8}$ is given by \cite{Andreas:1999ng,
Curio:1998bva, Blumenhagen:global02}
\begin{equation}
\chi_{E_8}=120\int_S(3\eta^2-27\eta c_1+62c_1^2).
\end{equation}
For the case of $G=SU(n)$, the characteristic $\chi_{SU(n)}$ is
given by\footnote{Eqs.~(\ref{refined chi})-(\ref{chiSU5})
initially were derived in heterotic string compactifications
\cite{Curio:1998bva,Andreas:1999ng}. A priori, these formulae are
valid only for F-theory models with a heterotic dual. It was
observed in \cite{Blumenhagen:global02} that these formulae also
hold for some F-theory models which do not admit a heterotic dual.
However, this match fails in other examples observed in
\cite{Chen:2010ts}. In these examples, extra gauge groups appear
in regions away from $S$ and cannot be described by spectral
covers. We assume that Eqs.~(\ref{refined chi})-(\ref{chiSU5})
hold for our models. }
\begin{equation}
\chi_{SU(n)}=\int_S(n^3-n)c_1^2+3n\eta(\eta-nc_1).\label{chiSU5}
\end{equation}
When $G$ splits into a product of two groups $G_1$ and $G_1$,
$\chi_{G}$ in Eq. (\ref{refined chi}) is then replaced by
$\chi^{(k)}_{G_1}+\chi^{(l)}_{G_2}$ in which $\eta$ is replaced by
the class $\eta^{(m)}$ in the spectral cover $\mathcal{C}^{(m)}$
for $m=k,l$. For the case of $(3,1)$ factorization, the refined
Euler characteristic is then calculated by
\begin{eqnarray}
{\chi}(X_4)&=&\chi^{\ast}(X_4)+\chi^{(a)}_{SU(3)}+\chi^{(b)}_{SU(1)}
-\chi_{E_8}\nonumber\\&=&\chi^{\ast}(X_4)+\int_S3[c_1(38c_1-21t-20\xi_1)+(3t^2+6t\xi_1+4\xi_1^2)]-\chi_{E_8}.\label{refined
chi3,1}
\end{eqnarray}
In the $(2,2)$ factorization, the refined Euler
characteristic\footnote{For the $(3,1)$ factorization,
$\eta^{(a)}=(\eta-c_1-\xi_1)$ and $\eta^{(b)}=(c_1+\xi_1)$. For
the $(2,2)$ factorization, $\eta^{(d_1)}=(\eta-2c_1-\xi_2)$ and
$\eta^{(d_2)}=(2c_1+\xi_2)$.} is
\begin{eqnarray}
{\chi}(X_4)&=&\chi^{\ast}(X_4)+\chi^{(d_1)}_{SU(2)}+\chi^{(d_2)}_{SU(2)}-\chi_{E_8}\nonumber\\&=&
\chi^{\ast}(X_4)+\int_S6[c_1(10c_1-6t-4\xi_2)+(t^2+2t\xi_2+2\xi_2^2)]-\chi_{E_8}.\label{refined
chi2,2}
\end{eqnarray}

\subsection{Cover flux Contribution}

It follows from Eqs.~(\ref{Tadpole cancellation}) and
(\ref{self-intersection Gamma}) that
\begin{equation}
N_{D3}=\frac{{\chi}(X_4)}{24}+\frac{1}{2}\Gamma^2.\label{Tadpole
cancellation with Gamma}
\end{equation}
In the previous subsection, we discussed the first term on the
right hand side of Eq.~(\ref{Tadpole cancellation with Gamma}). To
calculate $N_{D3}$, it is necessary to compute the
self-intersection $\Gamma^2$ of the universal cover flux $\Gamma$.
Recall that in section 4, the universal cover flux was defined by
\begin{equation}
\Gamma=\sum_{k}\Gamma^{(k)},
\end{equation}
where $\Gamma^{(k)}$ are cover fluxes satisfying the traceless
condition,
\begin{equation}
\sum_{k}p_{k\ast}\Gamma^{(k)}=0.
\end{equation}
In what follows, we will compute $\Gamma^2$ for both the $(3,1)$
and $(2,2)$ factorizations.

\subsubsection{ $(3,1)$ Factorization}

Recall that for the case of $(3,1)$ factorization, the universal
cover flux is given by
\begin{equation}
\Gamma=k_a \gamma^{(a)}_0+k_b\gamma^{(b)}_0+m_a \delta^{(a)} +m_b
\delta^{(b)} + \tilde{\rho}=\Gamma^{(a)}+\Gamma^{(b)},
\end{equation}
where $\Gamma^{(a)}$ and $\Gamma^{(b)}$ are
\begin{equation}
\Gamma^{(a)}=[\mathcal{C}^{(a)}]\cdot \left[(3k_a+m_a)\sigma
-\pi^{\ast}(k_a [a_3] +m_b[d_1]+\rho)\right]
\equiv[\mathcal{C}^{(a)}]\cdot[\widetilde{\mathcal{C}}^{(a)}],\label{dual
cover a}
\end{equation}
\begin{equation}
\Gamma^{(b)}=[\mathcal{C}^{(b)}]\cdot \left[(k_b+3m_b)\sigma
-\pi^{\ast}(k_b[d_1]+m_a[a_3]-3\rho)
\right]\equiv[\mathcal{C}^{(b)}]\cdot[\widetilde{\mathcal{C}}^{(b)}].\label{dual
cover b}
\end{equation}
Then the self-intersection of the cover flux $\Gamma$ is
calculated by \cite{Caltech:global03}
\begin{equation}
\Gamma^2=[\mathcal{C}^{(a)}]\cdot[\widetilde{\mathcal{C}}^{(a)}]\cdot[\widetilde{\mathcal{C}}^{(a)}]+[\mathcal{C}^{{(b)}}]\cdot[\widetilde{\mathcal{C}}^{(b)}]
\cdot[\widetilde{\mathcal{C}}^{(b)}].
\end{equation}
In the $(3,1)$ factorization,
$[\mathcal{C}^{(a)}]=3\sigma+\pi^{\ast}(\eta-c_1-\xi_1)$ and
$[\mathcal{C}^{(b)}]=\sigma+\pi^{\ast}(c_1+\xi_1)$. By
Eqs.~(\ref{dual cover a}) and (\ref{dual cover b}), one can obtain
\begin{eqnarray}
[\mathcal{C}^{{(a)}}]\cdot[\widetilde{\mathcal{C}}^{(a)}]\cdot[\widetilde{\mathcal{C}}^{(a)}]&=&-(3k_a+m_a)^2([a_3]\cdot_S
c_1)-k_a(3k_a+2m_a)[a_3]^2+3m_b^2[d_1]^2\nonumber\\&-&2m_bm_a([a_3]\cdot_S[d_1])-2(m_a[a_3]-3m_b[d_1])
\cdot_S\rho\nonumber\\&+&3(\rho\cdot_S\rho),
\end{eqnarray}
and
\begin{eqnarray}
[\mathcal{C}^{{(b)}}]\cdot[\widetilde{\mathcal{C}}^{(b)}]
\cdot[\widetilde{\mathcal{C}}^{(b)}]&=&- (k_b+3m_b)^2([d_1]\cdot_S
c_1)-k_b(k_b+6m_b)[d_1]^2+m_a^2[a_3]^2
\nonumber\\&-&6m_bm_a([a_3]\cdot_S[d_1])-6(m_a[a_3]-3m_b[d_1])\cdot_S\rho\nonumber\\&+&9(\rho\cdot_S\rho).
\end{eqnarray}
Putting everything together, one obtains
\begin{equation}
\Gamma^2=-\frac{1}{3}(3k_a+m_a)^2([a_0]\cdot_S
[a_3])-(k_b+3m_b)^2([d_0]\cdot_S
[d_1])+\frac{4}{3}(m_a[a_3]-3m_b[d_1]-3\rho)^2.
\end{equation}

\subsubsection{ $(2,2)$ Factorization}

Recall that in the $(2,2)$ factorization, the universal flux is
given by
\begin{equation}
\Gamma=k_{d_1}\gamma^{(d_1)}_0+k_{d_2}\gamma^{(d_2)}_0+m_{d_1}\delta^{(d_1)}+m_{d_2}\delta^{(d_2)}+\widehat{{\rho}}\equiv\Gamma^{(d_1)}+\Gamma^{(d_2)},
\end{equation}
where $\Gamma^{(d_1)}$ and $\Gamma^{(d_2)}$ are
\begin{equation}
\Gamma^{(d_1)}=[\mathcal{C}^{(d_1)}]\cdot[2(k_{d_1}+m_{d_1})\sigma-\pi^{\ast}(k_{d_1}[e_2]+m_{d_2}[f_2]+\rho)]\equiv[\mathcal{C}^{(d_1)}]
\cdot[\widetilde{\mathcal{C}}^{(d_1)}],\label{dual cover d_1}
\end{equation}
\begin{equation}
\Gamma^{(d_2)}=[\mathcal{C}^{(d_2)}]\cdot[2(k_{d_2}+m_{d_2})\sigma-\pi^{\ast}(k_{d_2}[f_2]+m_{d_1}[e_2]-\rho)]\equiv[\mathcal{C}^{(d_2)}]
\cdot[\widetilde{\mathcal{C}}^{(d_2)}].\label{dual cover d_2}
\end{equation}
Then the self-intersection $\Gamma^2$ can be computed as
\begin{equation}
\Gamma^2=
[\mathcal{C}^{{(d_1)}}]\cdot[\widetilde{\mathcal{C}}^{(d_1)}]\cdot[\widetilde{\mathcal{C}}^{(d_1)}]+[\mathcal{C}^{(d_2)}]\cdot[\widetilde{\mathcal{C}}^{{(d_2)}}]
\cdot[\widetilde{\mathcal{C}}^{(d_2)}].
\end{equation}
Notice that
$[\mathcal{C}^{(d_1)}]=2\sigma+\pi^{\ast}(\eta-2c_1-\xi_2)$ and
$[\mathcal{C}^{(d_2)}]=2\sigma+\pi^{\ast}(2c_1+\xi_2)$ in the
$(2,1)$ factorization. It follows from Eqs.~(\ref{dual cover d_1})
and (\ref{dual cover d_2}) that
\begin{eqnarray}
[\mathcal{C}^{{(d_1)}}]\cdot[\widetilde{\mathcal{C}}^{(d_1)}]\cdot[\widetilde{\mathcal{C}}^{(d_1)}]&=&
-4(k_{d_1}+m_{d_1})^2([e_2]\cdot_S
c_1)-2k_{d_1}(k_{d_1}+2m_{d_1})[e_2]^2+2m_{d_2}^2[f_2]^2\nonumber\\&-&4m_{d_1}m_{d_2}([e_2]\cdot_S
[f_2])-4(m_{d_1}[e_2]-m_{d_2}[f_2])\cdot_S\rho\nonumber\\&+&2(\rho\cdot_S\rho),
\end{eqnarray}
and
\begin{eqnarray}
[\mathcal{C}^{(d_2)}]\cdot[\widetilde{\mathcal{C}}^{{(d_2)}}]
\cdot[\widetilde{\mathcal{C}}^{(d_2)}]&=&-4(k_{d_2}+m_{d_2})^2([f_2]\cdot_S
c_1)-2k_{d_2}(k_{d_2}+2m_{d_2})[f_2]^2+2m_{d_1}^2[e_2]^2\nonumber\\&-&4m_{d_1}m_{d_2}([f_2]\cdot_S
[e_2])-4(m_{d_1}[e_2]-m_{d_2}[f_2])\cdot_S\rho\nonumber\\&+&2(\rho\cdot_S\rho).
\end{eqnarray}
Therefore, $\Gamma^2$ is given by
\begin{equation}
\Gamma^2=-2(k_{d_1}+m_{d_1})^2([e_0]\cdot_S
[e_2])-2(k_{d_2}+m_{d_2})^2([f_0]\cdot_S
[f_2])+4(m_{d_1}[e_2]-m_{d_2}[f_2]-\rho)^2.
\end{equation}
%\footnote{Let $[F]$ be a supersymmetric class. We have $[F]\cdot_S\omega=0=\int_{S} F\wedge
%J=\int_{S}F\wedge\ast J=\int_{S}\ast F\wedge J$ where $F$ and $J$
%are the dual 2-form and K\"ahler form of the classes $[F]$ and
%$\omega$, respectively. It follows that $\ast F=-F$ and then
%$[F]^2=\int_S F\wedge F=-\int_S F\wedge\ast F<0$}

\section{Models}

\subsection{$U(1)_X$ Flux and Spectrum}

Let us start with the $(3,1)$ factorization. Consider the breaking
pattern as follows:
\begin{equation}
\begin{array}{c@{}c@{}l@{}c@{}l}
SU(4)_{\bot} &~\rightarrow~& SU(3)\times U(1)\\
{\bf 15} &~\rightarrow~& {\bf 8}_0+{\bf 3}_{-4}+{\bf\bar 3}_4+{\bf 1}_0\\
{\bf 6} &~\rightarrow~& {\bf 3}_{2}+{\bf\bar 3}_{-2}\\
{\bf 4} &~\rightarrow~& {\bf 3}_{-1}+{\bf 1}_{3}\label{SU(3)U(1)}
\end{array}
\end{equation}
Then the representations $({\bf 16,4})$ and $({\bf 10},{\bf 6})$
in Eq. (\ref{E_8 decomposition}) are decomposed as
\begin{equation}
({\bf 16,4})\ra ({\bf 16}_{-1},{\bf 3})+({\bf 16}_{3},{\bf
1}),\;\;\; ({\bf 10},{\bf 6})\ra({\bf 10}_{2},{\bf 3})+ ({\bf
10}_{-2},{\bf\bar 3})\label{Tate form 2}
\end{equation}

\begin{table}[h]
\begin{center}
\renewcommand{\arraystretch}{1}
\begin{tabular}{|c|c|c|c|} \hline
Curve & Matter & Bundle & Chirality  \\ \hline

\multirow{3}{*}{${\bf 16}^{(a)}_{-1}$} & ${\bf 10}_{-1,-1}$
& $V_{{\bf 16}}\otimes L^{-1}_X|_{\Sigma^{(a)}_{\bf 16}}$ & $M_a$ \\

& $\bar{\bf 5}_{-1,3}$ & $V_{{\bf 16}}\otimes L^3_X|_{\Sigma^{(a)}_{\bf 16}}$ & $M_a+N_a$ \\

& ${\bf 1}_{-1,-5}$ & $V_{{\bf 16}}\otimes
L^{-5}_X|_{\Sigma^{(a)}_{\bf 16}}$ & $M_a-N_a$ \\ \hline

\multirow{3}{*}{${\bf 16}^{(b)}_{3}$} & ${\bf 10}_{3,-1}$
& $V_{{\bf 16}}\otimes L^{-1}_X|_{\Sigma^{(b)}_{\bf 16}}$ & $M_b$\\

& $\bar{\bf 5}_{3,3}$ & $V_{{\bf 16}}\otimes L^3_X|_{\Sigma^{(b)}_{\bf 16}}$ & $M_b+N_b$ \\

& ${\bf 1}_{3,-5}$ & $V_{{\bf 16}}\otimes
L^{-5}_X|_{\Sigma^{(b)}_{\bf 16}}$ & $M_b-N_b$  \\ \hline

\multirow{2}{*}{${\bf 10}^{(a)(a)}_{-2}$} & ${\bf 5}_{-2,2}$ &
$V_{{\bf 10}}\otimes L^2_X|_{\Sigma^{(a)(a)}_{\bf 10}}$ & $M_{aa}+N_{aa}$ \\

& $\bar{\bf 5}_{-2,-2}$ & $V_{{\bf 10}}\otimes
L^{-2}_X|_{\Sigma^{(a)(a)}_{\bf 10}}$ & $M_{aa}$ \\ \hline

\multirow{2}{*}{${\bf 10}^{(a)(b)}_2$} & ${\bf 5}_{2,2}$ &
$V_{{\bf 10}}\otimes L^2_X|_{\Sigma^{(a)(b)}_{\bf 10}}$ & $M_{ab}+N_{ab}$ \\

& $\bar{\bf 5}_{2,-2}$ & $V_{{\bf 10}}\otimes
L^{-2}_X|_{\Sigma^{(a)(b)}_{\bf 10}}$ & $M_{ab}$ \\ \hline
\end{tabular}
\caption{Chirality of matter localized on matter curves ${\bf 16}$
and ${\bf 10}$ in the (3,1) factorization.} \label{3-1FX}
\end{center}
\end{table}

%%%%%%%%%%%%%%%%%%%%%%%%%%%%%%%%%%%%%%%%%%%%%%%%%%%%%%%%%%%%%%%%%%%%%%%%%%%%%%%%%%%%%%%%%%%%%%%%%%%%%
On the other hand, we can further break $SO(10)$ in Eq. (\ref{E_8
decomposition}) by $U(1)_X$ flux as follows:
\begin{equation}
\begin{array}{c@{}c@{}l@{}c@{}l}
SO(10) &~\rightarrow~& SU(5)\times U(1)_{X}\\
{\bf 16} &~\rightarrow~& {\bf 10}_{-1}+{\bf\bar 5}_3+{\bf
1}_{-5}\\{\bf 10} &~\rightarrow~& {\bf 5}_{2}+{\bf\bar 5}_{-2}
\end{array}
\end{equation}
We suppose that $V_{\bf 16}\otimes L^{-1}_X$ has restriction of
degree $M_k$ to  $\Sigma_{{\bf 16}^{(k)}}$  while $L_X^4$ has
restriction of degree $N_k$. Similarly, we define $V_{\bf
10}\otimes L^{-2}_X$ has restriction of degree $M_{kl}$ to
$\Sigma_{{\bf 10}^{(k)(l)}}$ while $L_X^4$ has restriction of
degree $N_{kl}$. We summarize the chirality on each matter curve
in Table {\ref{3-1FX}}.  For the $(2,2)$ factorization, the
analysis is similar to the case of the $(3,1)$ factorization. We
summarize the chirality induced from the cover and $U(1)_X$ fluxes
in Table \ref{2-2FX}.
%\footnote{we define $\Gamma|_{\Sigma_{{\bf
%16}}^{(k)}}=M_k={\rm deg}(V_{{\bf 16}^{(k)}}\otimes L_X^{-1})$,
%$\Gamma|_{\Sigma_{{\bf 10}^{(k)(l)}}}=M_{kl}={\rm deg}(V_{{\bf
%10}^{(k)(l)}}\otimes L_X^{-2})$, $L^{4}_{X}|_{\Sigma_{{\bf
%16}}^{(k)}}=N_k$, and $L^{4}_{X}|_{\Sigma_{{\bf
%10}}^{(k)(l)}}=N_{kl}$.}

\begin{table}[h]
\begin{center}
\renewcommand{\arraystretch}{1}
\begin{tabular}{|c|c|c|c|} \hline
Curve & Matter & Bundle & Chirality  \\ \hline

\multirow{3}{*}{${\bf 16}^{(d_2)}_{-1}$} & ${\bf 10}_{-1,-1}$
& $V_{{\bf 16}}\otimes L^{-1}_X|_{\Sigma^{(d_2)}_{\bf 16}}$ & $M_{d_2}$\\

& $\bar{\bf 5}_{-1,3}$ & $V_{{\bf 16}}\otimes L^3_X|_{\Sigma^{(d_2)}_{\bf 16}}$ & $M_{d_2}+N_{d_2}$ \\

& ${\bf 1}_{-1,-5}$ & $V_{{\bf 16}}\otimes
L^{-5}_X|_{\Sigma^{(d_2)}_{\bf 16}}$ & $M_{d_2}-N_{d_2}$  \\\hline

\multirow{3}{*}{${\bf 16}^{(d_1)}_{1}$} & ${\bf 10}_{1,-1}$
& $V_{{\bf 16}}\otimes L^{-1}_X|_{\Sigma^{(d_1)}_{\bf 16}}$ & $M_{d_1}$ \\

& $\bar{\bf 5}_{1,3}$ & $V_{{\bf 16}}\otimes L^3_X|_{\Sigma^{(d_1)}_{\bf 16}}$ & $M_{d_1}+N_{d_1}$  \\

& ${\bf 1}_{1,-5}$ & $V_{{\bf 16}}\otimes
L^{-5}_X|_{\Sigma^{(d_1)}_{\bf 16}}$ & $M_{d_1}-N_{d_1}$  \\
\hline

\multirow{2}{*}{${\bf 10}^{(d_2)(d_2)}_{-2}$} & ${\bf 5}_{-2,2}$ &
$V_{{\bf 10}}\otimes L^2_X|_{\Sigma^{(d_2)(d_2)}_{\bf 10}}$ & $M_{d_2d_2}+N_{d_2d_2}$ \\
& $\bar{\bf 5}_{-2,-2}$ & $V_{{\bf 10}}\otimes
L^{-2}_X|_{\Sigma^{(d_2)(d_2)}_{\bf 10}}$ & $M_{d_2d_2}$ \\ \hline

\multirow{2}{*}{${\bf 10}^{(d_1)(d_2)}_0$} & ${\bf 5}_{0,2}$ &
$V_{{\bf 10}}\otimes L^2_X|_{\Sigma^{(d_1)(d_2)}_{\bf 10}}$ & $M_{d_1d_2}+N_{d_1d_2}$ \\

& $\bar{\bf 5}_{0,-2}$ & $V_{{\bf 10}} \otimes
L^{-2}_X|_{\Sigma^{(d_1)(d_2)}_{\bf 10}}$ & $M_{_{d_1d_2}}$
\\ \hline

\multirow{2}{*}{${\bf 10}^{(d_1)(d_1)}_{2}$} & ${\bf 5}_{2,2}$ &
$V_{{\bf 10}} \otimes L^2_X|_{\Sigma^{(d_1)(d_1)}_{\bf 10}}$ & $M_{d_1d_1}+N_{d_1d_1}$ \\

& $\bar{\bf 5}_{2,-2}$ & $V_{{\bf 10}}\otimes
L^{-2}_X|_{\Sigma^{(d_1)(d_1)}_{\bf 10}}$ & $M_{d_1d_1}$ \\ \hline
\end{tabular}
\caption{Chirality of matter localized on matter curves ${\bf 16}$
and ${\bf 10}$ in the (2,2) factorization.} \label{2-2FX}
\end{center}
\end{table}
%%%%%%%%%%%%%%%%%%%%%%%%%%%%%%%%%%%%%%%%%%%%%%%%%%%%%%%%%%%%%%%%%%%%%%%%%%%%%%%%%%%%%

%%%%%%%%%%%%%%%%%%%%%%%%%%%%%%%%%%%%%%%%
%%%%%%%%%%%%%%%%%%%%%%%%%%%%%%%%%%%%%%%%

\subsection{(3,1) Factorization and $CY_4$ with a $dP_2$ Surface}

In this section, we shall explicitly realize models in specific
geometries. We first consider the Calabi-Yau fourfold constructed
in \cite{Caltech:global01} to be our $X_4$. This Calabi-Yau
fourfold contains a $dP_2$ surface embedded into the base $B_3$.
For the detailed geometry of this Calabi-Yau fourfold, we refer
readers to \cite{Caltech:global01}. Here we only collect the
relevant geometric data\footnote{In section 6, $H$ and
$E_m,\;m=1,2,..,k$ are defined to be the hyperplane divisor and
exceptional divisors of $dP_k$, respectively.} for calculation.
The basic geometric data of $X_4$ is
\begin{equation}
c_1= 3H-E_1-E_2,~~
t=-c_1(N_{S/B_3})=H,~~\chi^{\ast}(X_4)=13968.\label{G data CIT
3,1}
\end{equation}
From Eq. (\ref{G data CIT 3,1}), we can conclude
$\eta=17H-6E_1-6E_2$, $\eta^2=217$, $c_1\cdot\eta=39$, and
$c_1^2=7$. For the (3,1) factorization, it follows from Eq.
(\ref{refined chi3,1}) that the refined Euler characteristic is
\begin{equation}
\chi(X_4)=10746+(12\xi_1^2 -18\xi_1\eta + 48\xi_1
c_1).\label{dp2chi31}
\end{equation}
The self-intersection of the cover flux $\Gamma$ is then given by
\begin{eqnarray}
\Gamma^2 &=& -(3k_a^2+2k_am_a)(50+\xi_1^2-2\xi_1\eta+5\xi_1 c_1)
+m_a^2(6+\xi_1^2-2\xi_1\eta+9\xi_1 c_1) \nonumber \\
&& -(k_b+3m_b)^2(\xi_1^2+\xi_1
c_1)+12m_b^2\xi_1^2+8m_am_b(\xi_1^2-\xi_1\eta+4\xi_1 c_1) \nonumber \\
&& +12\rho^2 -8m_a(\rho\eta-\rho\xi_1-4\rho
c_1)+24m_b\rho\xi_1,\label{gamma square 3,1}
\end{eqnarray}
and the number of generations for matter ${\bf 16}$ and ${\bf 10}$
on the curves are
\begin{eqnarray}
N_{{\bf 16}^{(b)}} &=& (m_a-k_b)\xi_1^2
-m_a\xi_1\eta+(4m_a-k_b-3m_b)\xi_1 c_1 + 3\rho\xi_1, \\
N_{{\bf 16}^{(a)}} &=&
-(50k_a+11m_a)+(m_b-k_a)\xi_1^2+(2k_a-m_b)\xi_1\eta \nonumber
\\&& +(4m_b-5k_a+m_a)\xi_1c_1  -\rho\eta+4\rho c_1+\rho\xi_1,  \\
N_{{\bf 10}^{(a)(b)}} &=& -28(k_a+m_a)-(k_b+3k_a+m_a+3m_b)\xi_1^2
+(4k_a+2m_a+2m_b)\xi_1\eta \nonumber \\
&&-(k_b+15k_a+7m_a+9m_b)\xi_1 c_1 + 2\rho\eta-6\rho c_1, \\
N_{{\bf 10}^{(a)(a)}} &=& 28(k_a+m_a)+(3k_a+m_a)\xi_1^2
-(4k_a+2m_a+2m_b)\xi_1\eta \nonumber \\ &&+(15k_a+7m_a+6m_b)\xi_1
c_1 - 2\rho\eta+6\rho c_1.
\end{eqnarray}
In this case, the supersymmetric condition Eq. (\ref{simplified
cover constraint 03}) reduces to
\begin{eqnarray}
[(3m_b+m_a)\xi_1-m_a(\eta-4c_1)+3\rho]\cdot_S[\omega],
\end{eqnarray}
where we choose
$[\omega]=\alpha(E_1+E_2)+\beta(H-E_1-E_2),\;2\alpha>\beta>\alpha>0$
to be an ample divisor in $dP_2$. In the (3,1) factorization, one
more constraint that we may impose is that the ramification of the
degree-one cover should be trivial. In other words, we impose the
following constraint:
\begin{equation}
(c_1+\xi_1)\cdot\xi_1=0.
\end{equation}
In what follows, we show three examples based on this geometry. We
find that there are only finite number of solutions for
parameters.

%%%%%%%%%%%%%%%%%%%%%%%%%%%%%%%%%%%%%%%%%%%%%%%%%%%%%%%%

%%%%%%%%%%%%%%%%%%%%%%%%%%%%%%%%%%%%%%%%%%%%%%%%%%%%%
\subsubsection{Model 1}

In this model we represent a three-generation example. The
numerical parameters are listed in Table \ref{pM-1C31}.
\begin{table}[h] %Model 1
\center
\begin{tabular}{|c|c|c|c|c||c||c|c|}
\hline $k_b$ & $k_a$ & $m_b$ & $m_a$ & $\rho$ & $\xi_1$ & $\alpha$
& $\beta$ \\ \hline -1.5 & -0.5 & -2 & 1 & $H+3E_1+E_2$ &
$E_2$ & 9 & 11 \\
\hline
\end{tabular}
\caption{Parameters of Model 1 of the (3,1) factorization in
$dP_2$.} \label{pM-1C31}
\end{table}
\\The matter content and the corresponding classes are listed in
Table \ref{M-1C31}. By using Eqs.~(\ref{dp2chi31}) and (\ref{gamma
square 3,1}), we obtain $\chi(X_4)=10674$ and $\Gamma^2=-159.5$.
It follows from Eq.~(\ref{Tadpole cancellation with Gamma}) that
$N_{D3}=365$.

\begin{table}[h]
\center
\begin{tabular}{|l|c|c|c|c|} \hline
Matter & Class in $S$ & Class with fixed $\xi_1$ & Generation &
Restr. of $[F_X]$ \\\hline

${\bf 16}^{(b)}$ & $\xi_1$ &  $E_2$ & 0 & 1 \\ \hline

${\bf 16}^{(a)}$ & $\eta-4c_1-\xi_1$  & $5H-2E_1-3E_2$ & 3 & $-1$
\\\hline %

${\bf 10}^{(a)(b)}$ & $\eta-3c_1$ & $8H-3E_1-3E_2$ & 14 & 0
\\\hline %

${\bf 10}^{(a)(a)}$ & $\eta-3c_1$ & $8H-3E_1-3E_2$ & $-14$ & 0
\\\hline %
\end{tabular}
\caption{Model 1 matter content with $[F_X]=E_1-E_2$. It is a
three-generation model with non-trivial flux restrictions. }
\label{M-1C31}
\end{table}

%\begin{equation}
%\chi(X_4)=24\cdot(444.75),\;\;\Gamma^2=-159.5,\;\;N_{D3}=365.
%\end{equation}

%%%%%%%%%%%%%%%%%%%%%%%%%%%%%%%%%%%%%%%%%%%%%%%%%%%%%
\subsubsection{Model 2}

Model $2$ is another example of a three-generation model with
$\chi(X_4)=10674$, $\Gamma^2=-159.5$, and $N_{D3}=365$. The
construction is similar to the model $1$. We list the numerical
parameters in Table \ref{pM-2C31}.
\begin{table}[h] %Model 4
\center
\begin{tabular}{|c|c|c|c|c||c||c|c|}
\hline $k_b$ & $k_a$ & $m_b$ & $m_a$ & $\rho$ & $\xi_1$ & $\alpha$
& $\beta$ \\ \hline -1.5 & 0.5 & -2 & -2 & $-4H+4E_1+5E_2$ &
$E_1$ & 9 & 11 \\
\hline
\end{tabular}
\caption{Parameters of Model 2 of the (3,1) factorization in
$dP_2$.} \label{pM-2C31}
\end{table}
\\ The matter content and the corresponding classes are shown in
Table \ref{M-2C31}.
\begin{table}[h]
\center
\begin{tabular}{|l|c|c|c|c|} \hline
Matter & Class in $S$ & Class with fixed $\xi_1$ & Generation &
Restr. of $[F_X]$ \\\hline

${\bf 16}^{(b)}$ & $\xi_1$ &  $E_1$ & 0 & 1 \\ \hline %
${\bf 16}^{(a)}$ & $\eta-4c_1-\xi_1$  & $5H-3E_1-2E_2$ & 3 & $-1$
\\\hline
${\bf 10}^{(a)(b)}$ & $\eta-3c_1$ & $8H-3E_1-3E_2$ & 14 & 0
\\\hline
${\bf 10}^{(a)(a)}$ & $\eta-3c_1$ & $8H-3E_1-3E_2$ & $-14$ & 0
\\\hline
\end{tabular}
\caption{Model 2 matter content with $[F_X]=E_1-E_2$.}
\label{M-2C31}
\end{table}

%%%%%%%%%%%%%%%%%%%%%%%%%%%%%%%%%%%%%%%%%%%%%%%%%%%%%
\subsubsection{Model 3}

Next we build a four-generation model in $SO(10)$. The reason why
we would like to discuss such a case is that the only choice for
the $U(1)_X$ flux on $dP_2$ is $[F_X]=\pm(E_1-E_2)$, and then the
restrictions of $[F_X]$ to the $\bf 16$ curves are always
non-zero, which results in the variation of the chirality numbers
of the $SU(5)$ matter descended from the $\bf 16$ curves. The two
examples shown above only make sense for an three-generation
$SO(10)$ model, and they are no longer three-generation models
after gauge breaking. Since we expect to build a three-generation
model at $SU(5)$ level, we slightly increase the generation number
at the $SO(10)$ level to prevent the chirality being too small.
The numerical parameters are listed in Table \ref{pM-3C31}. In
this model, it is not difficult to obtain $\chi(X_4)=10674$ and
$\Gamma^2=-355.5$. It turns out that $N_{D3}=267$ is a positive
integer.
\begin{table}[h] %Model 1 4gen
\center
\begin{tabular}{|c|c|c|c|c||c||c|c|}
\hline $k_b$ & $k_a$ & $m_b$ & $m_a$ & $\rho$ & $\xi_1$ & $\alpha$
& $\beta$ \\ \hline -1.5 & -0.5 & -2 & 1 & $5E_1+E_2$ &
$E_2$ & 12 & 17 \\
\hline
\end{tabular}
\caption{Parameters of Model 3 of the (3,1) factorization in
$dP_2$.} \label{pM-3C31}
\end{table}
\\The matter content and the corresponding classes are listed in
Table \ref{M-3C31}.
\begin{table}[h]
\center
\begin{tabular}{|l|c|c|c|c|} \hline
Matter & Class in $S$ & Class with fixed $\xi_1$ & Generation &
Restr. of $[F_X]$ \\\hline

${\bf 16}^{(b)}$ & $\xi_1$ &  $E_2$ & 0 & 1 \\ \hline

${\bf 16}^{(a)}$ & $\eta-4c_1-\xi_1$  & $5H-2E_1-3E_2$ & 4 & $-1$
\\\hline

${\bf 10}^{(a)(b)}$ & $\eta-3c_1$ & $8H-3E_1-3E_2$ & 10 & 0
\\\hline

${\bf 10}^{(a)(a)}$ & $\eta-3c_1$ & $8H-3E_1-3E_2$ & $-10$ & 0
\\\hline

\end{tabular}
\caption{Model 3 matter content with $[F_X]=E_1-E_2$. There are
four generations on the ${\bf 16}^{(a)}$ curve.} \label{M-3C31}
\end{table}

%%%%%%%%%%%%%%%%%%%%%%%%%%%%%%

\subsubsection{Discussion}

Model 1 and Model 2 of (3,1) factorization have the following
$SO(10)$ structure:
\begin{equation}
\begin{tabular}{c|c|c}
Maatter & Copy & $U(1)_C$ \\ \hline%
${\bf 16}^{(b)}$ & 0 & $-3$ \\
${\bf 16}^{(a)}$ & 3 & 1 \\
${\bf 10}^{(a)(b)}$ & $14$ & $-2$ \\
${\bf 10}^{(a)(a)}$ & $-14$ & 2 \\
\end{tabular}
\end{equation}
where $U(1)_C$ is from the cover, the $U(1)^3$ Cartan subalgebra
of $SU(4)_{\bot}$ that is not removed from the monodromy. The
Yukawa coupling is filtered by the conservation of this $U(1)_C$.
Before turning on the $U(1)_X$ flux, this spectrum can fit the
minimum requirement by forming the Yukawa coupling ${\bf
16}^{(a)}_{-1} {\bf 16}^{(a)}_{-1} {\bf 10}^{(a)(b)}_{2}$ of the
$SO(10)$ GUT with some exotic $\bf 10$s. However, when $U(1)_X$
flux is turned on, the non-vanishing restriction of the flux to
each $\bf 16$ curve changes the chirality, while the chirality on
the $\bf 10$ curves remain untouched. The analysis in Table
\ref{3-1FX} suggests that a three-generation model may descend
from a four-generation $SO(10)$ model after the gauge group is
broken to $SU(5)\times U(1)_X$ by $[F_X]=E_1-E_2$. Here we try to
explain Model 3 as a flipped $SU(5)$ model with its spectrum
presented in Table \ref{31dP2FSU5}.
\begin{table}[h]
\center
\begin{tabular}{|c|c|c|} \hline
Matter & Rep. & Generation \\ \hline

${\bf 10}_M$ & ${\bf 10}_{-1,-1}$ & 3 \\
$\bar{\bf 5}_M$ & $\bar{\bf 5}_{-1,3}$ & 3 \\
${\bf 1}_M$ & ${\bf 1}_{-1,-5}$ & 3  \\ \hline

${\bf 10}_H+\overline{\bf 10}_H$ & ${\bf 10}_{-1,-1}+\overline{\bf 10}_{-1,1}$ & 1 \\
\hline

${\bf 5}_h$ & ${\bf 5}_{2,2}$ & 1 \\
$\bar{\bf 5}_h$ & $\bar{\bf 5}_{2,-2}$ & 1 \\  \hline \hline

${\bf 10}$ & ${\bf 10}_{-1,-1}$ & 1 \\
$\bar{\bf 5}$ &  $\bar{\bf 5}_{3,3}$ & 1 \\
${\bf 1}$ &  ${\bf 1}_{-1,-5}$ & 2 \\
${\bf 1}$ &  ${\bf 1}_{3,5}$ & 1 \\ \hline

\multirow{2}{*}{${\bf 5}+\bar{\bf 5}$ exotics} & ${\bf
5}_{-2,2}+\bar{\bf
5}_{-2,-2}$ & 9 \\

& ${\bf 5}_{2,2}+\bar{\bf 5}_{2,-2}$  & -10
\\ \hline
\end{tabular}
\caption{Flipped $SU(5)$ spectrum of Model 3.} \label{31dP2FSU5}
\end{table}

In this case, the Yukawa couplings are
\begin{eqnarray}
\mathcal{W}&\supset&{\bf 10}_{-1,-1M} {\bf 10}_{-1,-1M} {\bf
5}_{2,2h} + {\bf 10}_{-1,-1M} \bar{\bf 5}_{-1,3M}\bar{\bf
5}_{2,-2h} +
\bar{\bf 5}_{-1,3M} {\bf 1}_{-1,-5M} {\bf 5}_{2,2h} \nonumber \\
&+&  {\bf 10}_{-1,-1H} {\bf 10}_{-1,-1H} {\bf 5}_{2,2h} +
\overline{\bf 10}_{-1,1H} \overline{\bf 10}_{-1,1H} \bar{\bf
5}_{2,-2h}+\dots.
\end{eqnarray}
We may identify the flipped $SU(5)$ superheavy Higgs fields with
one of the ${\bf 10}+\overline{\bf 10}$ vector-like pairs on the
${\bf 16}^{(a)}$ curve, which is not obvious from this
configuration. Since the restrictions of the flux to the curves
change the chirality, there are unavoidable exotic fermions, like
the examples studied in \cite{Caltech:global03}. In the following
subsection, we will study models from a different geometric
backgrounds to see if it is possible to retain the chirality
unchanged while the flux $F_X$ is turned on.

\subsection{(3,1) Factorization and $CY_4$ with a $dP_7$ Surface}

Although $dP_2$ surface is elegant, it does not possess enough
degrees of freedom in the number of exceptional divisors for model
building. Therefore, we turn to the geometry of the compact
Calabi-Yau fourfold realized as complete intersections of two
hypersurfaces with an embedded $dP_7$ surface\footnote{By abuse of
notation, we also denote this Calabi-Yau fourfold by $X_4$.}. The
detailed construction can be found in\cite{Blumenhagen:global02}.
Again here we only collect relevant geometric data for
calculation. The basic geometric data is as follows:
\begin{eqnarray}
c_1&=& 3H-E_1-E_2-E_3-E_4-E_5-E_6-E_7, \nonumber \\
t &=& 2H-E_1-E_2-E_3-E_4-E_5-E_6, \nonumber\\
 \eta&=&16H-5E_1-5E_2-5E_3-5E_4-5E_5-5E_6-6E_7.\label{G data German 3,1}
\end{eqnarray}
with $\chi^{\ast}(X_4)=1728$. From Eq. (\ref{G data German 3,1}),
we have $\eta^2=70$, $\eta\cdot c_1=12$, and $c_1^2=2$. The
refined Euler characteristic is given by
\begin{equation}
\chi(X_4)=738+(12\xi_1^2 -18\xi_1\eta + 48\xi_1
c_1),\label{dp7chi31}
\end{equation}
and the self-intersection of the cover flux $\Gamma$ is
\begin{eqnarray}
\Gamma^2 &=& -(3k_a^2+2k_am_a)(18+\xi_1^2-2\xi_1\eta+5\xi_1 c_1)
+m_a^2(2
+\xi_1^2-2\xi_1\eta+9\xi_1 c_1) \nonumber \\
&& -(k_b+3m_b)^2(\xi_1^2+\xi_1
c_1)+12m_b^2\xi_1^2+8m_am_b(\xi_1^2-\xi_1\eta+4\xi_1 c_1) \nonumber \\
&& +12\rho^2 -8m_a(\rho\eta-\rho\xi_1-4\rho
c_1)+24m_b\rho\xi_1.\label{gamma square 3,1 dp7}
\end{eqnarray}
Again we summarize the generation number on each curve as follows:
\begin{eqnarray}
N_{{\bf 16}^{(b)}} &=& (m_a-k_b)\xi_1^2
-m_a\xi_1\eta+(4m_a-k_b-3m_b)\xi_1 c_1 + 3\rho\xi_1, \\
N_{{\bf 16}^{(a)}} &=& -(18k_a+4m_a)+(m_b-k_a)\xi_1^2+
(2k_a-m_b)\xi_1\eta \nonumber \\&& +(4m_b-5k_a+m_a)\xi_1
c_1 -\rho\eta+4\rho c_1+\rho\xi_1,  \\
N_{{\bf 10}^{(a)(b)}} &=& -10(k_a+m_a)-(k_b+3k_a+m_a+3m_b)\xi_1^2
+(4k_a+2m_a+2m_b)\xi_1\eta \nonumber \\
&& -(k_b+15k_a+7m_a+9m_b)\xi_1
c_1+ 2\rho\eta-6\rho c_1, \\
N_{{\bf 10}^{(a)(a)}} &=&10 (k_a+m_a)+(3k_a+m_a)\xi_1^2
-(4k_a+2m_a+2m_b)\xi_1\eta \nonumber \\ &&+(15k_a+7m_a+6m_b)\xi_1
c_1 - 2\rho\eta+6\rho c_1.
\end{eqnarray}
The supersymmetry condition is then
\begin{eqnarray}
[(3m_{b}+m_a)\xi_1 -m_{a}(\eta-4c_1)+3\rho]\cdot_S[\omega]=0,
\end{eqnarray}
where $[\omega]$ is an ample divisor dual to a K\"ahler form of
$dP_7$. For simplicity, we choose $[\omega]$ to be
\begin{eqnarray}
[\omega]=14\beta H-(5\beta-\alpha)\sum_{i=1}^7E_i,
\end{eqnarray}
with constraints $5\beta>\alpha>0$.

In what follows, we present one example based on this geometry.
This model is three-generation with vanishing restrictions of the
$U(1)_X$ flux to the $\bf 16$ curves.

%%%%%%%%%%%%%%%%%%%%%%%%%%%%%%%%%%%%%%%%%%%%%%%%%%%%%
\subsubsection{Model}

We present a three-generation model in this example. The numerical
result of the parameters is listed in Table \ref{pM-1G31}. With
data in Table\ref{pM-1G31} and Table\ref{M-1G31}, one can obtain
$\chi(X_4)=648$ and $\Gamma^2=-42$ by using Eqs.~(\ref{dp7chi31})
and (\ref{gamma square 3,1 dp7}). It follows from
Eq.~(\ref{Tadpole cancellation with Gamma}) that $N_{D3}=6$.
\begin{table}[h] %Model 100 FSOG31n
\center
\begin{tabular}{|c|c|c|c|c||c||c|c|}
\hline $k_b$ & $k_a$ & $m_b$ & $m_a$ & $\rho$ & $\xi_1$ & $\alpha$
& $\beta$ \\ \hline -1.5 & -1 & 0 & 1.5 & $\frac{1}{2}
(2E_1+2E_2+E_4)$ &
$2H-E_1-E_2-E_3-E_5-E_6$ & 3 & 2 \\
\hline
\end{tabular}
\caption{Parameters of the (3,1) factorization model in $dP_7$.}
\label{pM-1G31}
\end{table}
\\ The matter content and the corresponding classes are listed in
Table \ref{M-1G31}.
\begin{table}[h]
\center
\begin{tabular}{|l|c|c|c|c|} \hline
Matter & Class in $S$ & Class with fixed $\xi_1$ & Generation
\\\hline

${\bf 16}^{(b)}$ & $\xi_1$ & $2H-E_1-E_2-E_3-E_5-E_6$ & 0 \\ %$H-E_1-E_2+E_4+E_5+E_6$ & 0 \\
\hline

${\bf 16}^{(a)}$ & $\eta-4c_1-\xi_1$  & $2H-E_4-2E_7$ & 3
\\\hline

${\bf 10}^{(a)(b)}$ & $\eta-3c_1$ & $7H-2\sum_{i=1}^6E_i-3E_7$ & 1
\\\hline

${\bf 10}^{(a)(a)}$ & $\eta-3c_1$ & $7H-2\sum_{i=1}^6E_i-3E_7$ &
-1
\\\hline

\end{tabular}
\caption{The $dP_7$ model matter content. Since it is a
three-generation model, the flux is chosen to have trivial
restriction. For example, $[F_X]=E_5-E_6$.} \label{M-1G31}
\end{table}

%%%%%%%%%%%%%%%%%%%%%%%%%%%%%%%%%%%%%%%%%%%%%%%%%%%%%
%\subsubsection{Model 2}

%The numerical parameters are listed in Table \ref{pM-2G31}.
%\begin{table}[h] %Model 350
%\center
%\begin{tabular}{|c|c|c|c|c||c||c|c|}
%\hline $k_b$ & $k_a$ & $m_b$ & $m_a$ & $\rho$ & $\xi_1$ & $\alpha$
%& $\beta$ \\ \hline 1.5 & 1.5 & 0 & -1 & $-H+E_1+E_2-E_7$ &
%$E_1+E_2+E_5$ & 13 & 4 \\
%\hline
%\end{tabular}
%\caption{Parameters of Model 2 of $dP_7$. $k_a$, $k_b$, $m_a$,
%$m_b$ and $\rho$ are the parameters of the flux, and $\alpha$ and
%$\beta$ are the parameters from the supersymmetry condition.}
%\label{pM-2G31}
%\end{table}
%\\ The matter content and the corresponding classes are listed in Table
%\ref{M-2G31}.
%\begin{table}[h]
%\center
%\begin{tabular}{|l|c|c|c|} \hline
%Matter & Class in $S$ & Class with fixed $\xi_1$ & Generation
%\\\hline

%${\bf 16}^{(b)}$ & $\xi_1$ &  $E_1+E_2+E_5$ & 0 \\
%\hline

%${\bf 16}^{(a)}$ & $\eta-4c_1-\xi_1$  &
%$4H-2E_1-2E_2-E_3-E_4-2E_5-E_6-2E_7$ & 3 \\\hline

%${\bf 10}^{(a)(b)}$ & $\eta-3c_1$ & $7H-2\sum_i^6E_i-3E_7$ & 7
%\\\hline

%${\bf 10}^{(a)(a)}$ & $\eta-3c_1$ & $7H-2\sum_i^6E_i-3E_7$ & -7
%\\\hline

%\end{tabular}
%\caption{Model 2 matter content. The flux is chosen to have
%trivial restriction. For example, $[F_X]=E_1-E_2$.}\label{M-2G31}
%\end{table}

\subsubsection{Discussion}

In this example we tune $[F_X]=E_4-E_5$ to obtain trivial
restrictions on all the curves, so the chirality on each curve
remains unchanged. By the analysis of Table \ref{3-1FX}, we can
create a flipped $SU(5)$ spectrum as shown in Table
\ref{fsu522-f0}. The Yukawa couplings turn out to be
\begin{eqnarray}
\mathcal{W}&\supset&{\bf 10}_{-1,-1M} {\bf 10}_{-1,-1M} {\bf
5}_{2,2h} + {\bf 10}_{-1,-1M} \bar{\bf 5}_{-1,3M}\bar{\bf
5}_{2,-2h} +
\bar{\bf 5}_{-1,3M} {\bf 1}_{-1,-5M} {\bf 5}_{2,2h} \nonumber \\
&+&  {\bf 10}_{-1,-1H} {\bf 10}_{-1,-1H} {\bf 5}_{2,2h} +
\overline{\bf 10}_{-1,1H} \overline{\bf 10}_{-1,1H} \bar{\bf
5}_{2,-2h}+\cdots.
\end{eqnarray}

{\renewcommand{\thefootnote}{\fnsymbol{footnote}}

\begin{table}[h]
\center
\begin{tabular}{|c|c|c|} \hline
Matter & Rep. & Generation \\ \hline

${\bf 10}_M$ & ${\bf 10}_{-1,-1}$ & 3 \\
$\bar{\bf 5}_M$ & $\bar{\bf 5}_{-1,3}$ & 3 \\
${\bf 1}_M$ & ${\bf 1}_{-1,-5}$ & 3 \\ \hline

${\bf 5}_h$ & ${\bf 5}_{2,2}$ & 1 \\
$\bar{\bf 5}_h$ & $\bar{\bf 5}_{2,-2}$ & 1 \\\hline

${\bf 10}_H+\overline{\bf 10}_H$ & ${\bf 10}_{-1,-1}+\overline{\bf 10}_{-1,1}$ & 1 \\
\hline \hline

\multicolumn{3}{|c|}{${\bf 5}+\bar{\bf 5}$ exotics\footnotemark[1]} \\
\hline

\end{tabular}
\caption{Flipped $SU(5)$ spectrum with vanishing restrictions of
$[F_X]$ on the curves in (3,1) factorization in $dP_7$.}
\label{fsu522-f0}
\end{table}
\footnotetext[1]{There is one $({\bf 5},\bar{\bf 5})$ on the ${\bf
10}^{(a)(a)}$ curve.} % in Model 1, and six $({\bf 5},\bar{\bf 5})$
%on ${\bf 10}^{(a)(b)}$ curve and seven $({\bf 5},\bar{\bf 5})$ on
%${\bf 10}^{(a)(a)}$ curve in Model 2.}

This spectrum looks standard, and the advantage is that there are
no exotic fermions and the quantum numbers(charges) of the matter
are typical. We again assume that the superheavy Higgses ${\bf
10}_H$ and $\overline{\bf 10}_H$ come from one of the vector-like
${\bf 10}+{\bf \overline{10}}$ pairs on the ${\bf 16}^{(a)}$
curve. It is not obvious to calculate the number of such pairs.
For simplicity, we just extract one pair for phenomenology
purposes.

%%%%%%%%%%%%%%%%%%%%%%%%%%%%%%%%%%%%%%%%%%%%%%%%
%%%%%%%%%%%%%%%%%%%%%%%%%%%%%%%%%%%%%%%%%%%%%%%%

\subsection{(2,2) Factorization and $CY_4$ with a $dP_2$ Surface}

Let us consider the $(2,2)$ factorization with the geometric
background in Eq. (\ref{G data CIT 3,1})\cite{Caltech:global01}.
In this case, the refined Euler characteristic turns out to be
\begin{equation}
\chi(X_4)=10446+(12\xi_2^2 -12\xi_2\eta + 48\xi_2
c_1).\label{dp7chi22}
\end{equation}
The self-intersection of the cover flux $\Gamma$ is
\begin{eqnarray}
\Gamma^2 &=& -2(k_{d_1}+m_{d_1})^2(39+\xi_2^2-2\xi_2\eta+6\xi_2
c_1)
+4m_{d_1}^2(17+\xi_2^2-2\xi_2\eta+8\xi_2 c_1) \nonumber \\
&& -2(k_{d_2}+m_{d_2})^2(\xi_2^2+2\xi_2
c_1)+4m_{d_2}^2\xi_2^2+8m_{d_1}m_{d_2}(\xi_2^2-\xi_2\eta+4\xi_2 c_1) \nonumber \\
&& +4\rho^2 -8m_{d_1}(\rho\eta-\rho\xi_2-4\rho
c_1)+8m_{d_2}\rho\xi_2.\label{gamma square 2,2 dp7}
\end{eqnarray}
In this case, we can find models with integral $N_{D3}$. However,
to have more degrees of freedom for model building, we shall focus
on the geometry of the $CY_4$ with an embedded $dP_7$ surface
\cite{Blumenhagen:global02} in the next subsection.

\subsection{(2,2) Factorization and $CY_4$ with a $dP_7$ Surface}

We again consider the geometric background in Eq. (\ref{G data
German 3,1})and the (2,2) factorization. In this case, the refined
Euler characteristic is given by
\begin{eqnarray}
\chi(X_4)= 636+(12\xi_2^2-12\xi_2\eta+48\xi_2 c_1).
\end{eqnarray}
The self-intersection of the cover flux $\Gamma$ is
\begin{eqnarray}
\Gamma^2 &=& -2(k_{d_1}+m_{d_1})^2(14+\xi_2^2-2\xi_2\eta+6\xi_2
c_1)
+4m_{d_1}^2(6+\xi_2^2-2\xi_2\eta+8\xi_2 c_1) \nonumber \\
&& -2(k_{d_2}+m_{d_2})^2(\xi_2^2+2\xi_2
c_1)+4m_{d_2}^2\xi_2^2+8m_{d_1}m_{d_2}(\xi_2^2-\xi_2\eta+4\xi_2 c_1) \nonumber \\
&& +4\rho^2 -8m_{d_1}(\rho\eta-\rho\xi_2-4\rho
c_1)+8m_{d_2}\rho\xi_2.
\end{eqnarray}
The generations of matter on the curves are
\begin{eqnarray}
N_{{\bf 16}^{(d_2)}} &=& (m_{d_1}-k_{d_2})\xi_2^2
-m_{d_1}\xi_2\eta+(4m_{d_1}-2k_{d_2}-2m_{d_2})\xi_2 c_1 + \rho\xi_2, \\
N_{{\bf 16}^{(d_1)}} &=&
-(14k_{d_1}+8m_{d_1})+(m_{d_2}-k_{d_1})\xi_2^2+(2k_{d_1}-m_{d_2})\xi_2\eta
\nonumber \\&& +(4m_{d_2}-6k_{d_1}+2m_{d_1})\xi_2 c_1  -\rho\eta+4\rho c_1+\rho\xi_2,  \\
N_{{\bf 10}^{(d_2)(d_2)}} &=& -8m_{d_1}+2(m_{d_1}+m_{d_2})\xi_2^2
+ 2(m_{d_2}+5m_{d_1})\xi_2 c_1-2m_{d_1}\xi_2\eta \nonumber \\
&&+2\rho c_1+2\rho\xi_2, \\
N_{{\bf 10}^{(d_1)(d_2)}} &=& -2(k_{d_1}+m_{d_1})(6+2\xi^2_2-3\xi_2\eta+12\xi_2 c_1), \\
N_{{\bf 10}^{(d_1)(d_1)}} &=& (12k_{d_1}+20m_{d_1})
+(4k_{d_1}+2m_{d_1}-2m_{d_2})\xi_2^2 -2(3k_{d_1}+2m_{d_1})\xi_2\eta \nonumber \\
&&+(24k_{d_1}-2m_{d_2}+14m_{d_1})\xi_2 c_1-2\rho c_1 -2\rho\xi_2.
\end{eqnarray}
The supersymmetry condition is then
\begin{eqnarray}
[2m_{d_2}\xi_2 -2m_{d_1}(\eta-4c_1-\xi_2)+2\rho]\cdot_S[\omega]=0,
\end{eqnarray}
where $[\omega]$ is an ample divisor dual to a K\"ahler form of
$dP_7$. For simplicity, we choose $[\omega]$ to be
\begin{eqnarray}
[\omega]=14\beta H-(5\beta-\alpha)\sum_{i=1}^7E_i,
\end{eqnarray}
with constraints $5\beta>\alpha>0$.

In the (2,2) factorization of the $SU(4)$ cover, we expect the
matter spectrum for an $SO(10)$ model as
\begin{equation}
\begin{tabular}{l|c|c}
Maatter & Copy & $U(1)_{C}$ \\ \hline%
${\bf 16}^{(d_2)}$ & 0/3 & -1 \\
${\bf 16}^{(d_1)}$ & 3/0 & 1 \\
${\bf 10}^{(d_2)(d_2)}$ & $n_1$ & -2 \\
${\bf 10}^{(d_1)(d_2)}$ & $n_2$ & 0 \\
${\bf 10}^{(d_1)(d_1)}$ & $n_3$ & 2 \\
\end{tabular}
\end{equation}
The $U(1)_C$ is of the $U(1)^3$ Cartan subalgebra of
$SU(4)_{\bot}$ that is not removed from the monodromy. The Yukawa
coupling is filtered by the conservation of this $U(1)_C$. The
possible Yukawa couplings for constructing a minimum $SO(10)$ GUT
are then ${\bf 16}^{(d_1)} {\bf 16}^{(d_1)} {\bf 10}^{(d_2)(d_2)}$
and ${\bf 16}^{(d_2)} {\bf 16}^{(d_2)} {\bf 10}^{(d_1)(d_1)}$.  We
will demonstrate examples of the flipped $SU(5)$ GUT model from
the following models.

%%%%%%%%%%%%%%%%%%%%%%%%%%%%%
\subsubsection{Model 1}

In this example we demonstrate a three-generation model.  The
numerical parameters are shown in Table \ref{pM-1G22}, and the
matter content and the corresponding classes with the flux
$[F_X]=E_2-E_3$ are listed in Table \ref{pM-1G220}. By using
Eqs.~(\ref{dp7chi22}) and (\ref{gamma square 2,2 dp7}), we obtain
$\chi(X_4)=600$ and $\Gamma^2=-18$ which gives rise to
$N_{D3}=16$.
\begin{table}[h]%Model 5
\center
\begin{tabular}{|c|c|c|c|c||c||c|c|}
\hline $k_{d_2}$ & $k_{d_1}$ & $m_{d_2}$ & $m_{d_1}$ & $\rho$ &
$\xi_2$ & $\alpha$ & $\beta$                           \\ \hline
-1 & 0 & 1.5 & -0.5 & $-\frac{1}{2}(H-2E_1+2E_2+2E_3+2E_4+E_7)$ &
$H-E_1$ & 1 & 3 \\ \hline
\end{tabular}
\caption{Parameters of Model 1 of the (2,2) Factorization in
$dP_7$. } \label{pM-1G22}
\end{table}
\begin{table}[h]
\begin{tabular}{|l|c|c|c|c|} \hline
Matter & Class in $S$ & Class with fixed $\xi_2$ & Generation &
Restr. of $F_X$\\\hline

${\bf 16}^{(d_2)}$ & $\xi_2$ & $H-E_1$ & 0 & 0 \\ \hline

${\bf 16}^{(d_1)}$ & $\eta-4c_1-\xi_2$ & $3H-\sum_{i=2}^6
E_i-2E_7$ & 3 & 0 \\\hline

${\bf 10}^{(d_2)(d_2)}$ & $c_1+\xi_2$ & $4H-2E_1-\sum_{i=2}^6
E_i-2E_7$ & 4 & 0 \\\hline

${\bf 10}^{(d_1)(d_2)}$ & $2\eta-8c_1-2\xi_2$ & $6H-2\sum_{i=2}^6
E_i-4E_7$ & -3 & 0 \\\hline

${\bf 10}^{(d_1)(d_1)}$ & $c_1+\xi_2$ & $4H-2E_1-\sum_{i=2}^6
E_i-2E_7$ & -1 & 0 \\\hline
\end{tabular}
\caption{The Matter content of Model 1. The flux is tuned that the
restriction is zero on each curve.}\label{pM-1G220}
\end{table}

%%%%%%%%%%%%%%%%%%%%%%%%%%%%%%%%%%%%%%%%%%
\subsubsection{Model 2}

\begin{table}[h] % Model 1096  4gen
\center
\begin{tabular}{|c|c|c|c|c||c||c|c|} \hline %
$k_{d_2}$ & $k_{d_1}$ & $m_{d_2}$ & $m_{d_1}$ & $\rho$ &
$\xi_2$ & $\alpha$ & $\beta$\\ \hline %
1 & 0 & -0.5 & -0.5 & $-\frac{1}{2}(H-2E_1+2E_2-2E_3-E_7)$ &
$2H-E_1-E_2-E_3-E_7$ & 1 & 3 \\ \hline
\end{tabular}
\caption{Parameters of Model 2 of the (2,2) Factorization in
$dP_7$. } \label{pM-2G22}
\end{table}

In this model, we show a four-generation example with non-zero
restrictions of $F_X$ on the matter curves. The spectrum can
maintain a three-generation model after the gauge is broken to
$SU(5)\times U(1)_X$ by $F_X$. The parameters are presented in
Table \ref{pM-2G22}, while the matter content and the
corresponding classes with the flux $[F_X]=E_3-E_4$ are listed in
Table \ref{m-2G22}. In this model, we have $\chi(X_4)=600$ and
$\Gamma^2=-26$ which gives rise to $N_{D3}=12$.

\begin{table}[h]
\begin{tabular}{|l|c|c|c|c|} \hline
Matter & Class in $S$ & Class with fixed $\xi_2$ & Gen. & Restr.
of $F_X$ \\\hline

${\bf 16}^{(d_2)}$ & $\xi_2$ & $2H-E_1-E_2-E_3-E_7$ & 0 & 1 \\
\hline

${\bf 16}^{(d_1)}$ & $\eta-4c_1-\xi_2$ & $2H-E_4-E_5-E_6-E_7$ & 4
& -1 \\\hline

${\bf 10}^{(d_2)(d_2)}$ & $c_1+\xi_2$ &
$5H-2E_1-2E_2-2E_3-\sum_{i=4}^6 E_i-2E_7$ & 4 & 1 \\\hline

${\bf 10}^{(d_1)(d_2)}$ & $2\eta-8c_1-2\xi_2$ &
$4H-2E_4-2E_5-2E_6-2E_7$ & -3 & -2 \\ \hline

${\bf 10}^{(d_1)(d_1)}$ & $c_1+\xi_2$ &
$5H-2E_1-2E_2-2E_3-\sum_{i=4}^6 E_i-2E_7$ & -1 & 1
\\\hline
\end{tabular}
\caption{Matter content of Model 2. The flux $[F_X]=E_3-E_4$ has
restrictions on the curves.} \label{m-2G22}
\end{table}

\subsubsection{Discussion}

\begin{table}[h]
\center
\begin{tabular}{|c|c|c|} \hline
Matter & Rep. & Generation \\ \hline

${\bf 10}_M$ & ${\bf 10}_{1,-1}$ & 3 \\
$\bar{\bf 5}_M$ & $\bar{\bf 5}_{1,3}$ & 3 \\
${\bf 1}_M$ & ${\bf 1}_{1,-5}$ & 3 \\ \hline

${\bf 5}_h$ & ${\bf 5}_{-2,2}$ & 1 \\
$\bar{\bf 5}_h$ & $\bar{\bf 5}_{-2,-2}$ & 1 \\\hline

${\bf 10}_H+\overline{\bf 10}_H$ & ${\bf 10}_{1,-1}+\overline{\bf 10}_{1,1}$ & 1 \\
\hline \hline

\multirow{3}{*}{${\bf 5}+\bar{\bf 5}$ exotics} & ${\bf 5}_{-2,2}+\bar{\bf 5}_{-2,-2}$ & 3 \\
& ${\bf 5}_{0,2}+\bar{\bf 5}_{0,-2}$ & 3 \\
& ${\bf 5}_{2,2}+\bar{\bf 5}_{2,-2}$ & -1
\\ \hline

\end{tabular}
\caption{Flipped $SU(5)$ spectrum of Model 1 of the (2,2)
factorization in $dP_7$.} \label{22FSU5}
\end{table}

The number of $(-2)$ 2-cycles in $dP_7$ is large enough that it is
possible to remain the chirality unchanged by tuning $F_X$ with
vanishing restrictions on all the curves. An example is presented
in Model 1, and the corresponding flipped $SU(5)$ spectrum can be
found in Table \ref{22FSU5}.

The Yukawa couplings of the flipped $SU(5)$ model from Model 1
then are
\begin{eqnarray}
\mathcal{W}&\supset&{\bf 10}_{1,-1M} {\bf 10}_{1,-1M} {\bf
5}_{-2,2h} + {\bf 10}_{1,-1M} \bar{\bf 5}_{1,3M}\bar{\bf
5}_{-2,-2h} + \bar{\bf 5}_{1,3M} {\bf 1}_{1,-5M} {\bf 5}_{-2,2h}
\nonumber \\ &+&  {\bf 10}_{1,-1H} {\bf 10}_{1,-1H} {\bf
5}_{-2,2h} + \overline{\bf 10}_{1,1H} \overline{\bf 10}_{1,1H}
\bar{\bf 5}_{-2,-2h}+\dots.
\end{eqnarray}

Similar to the examples with trivial restriction of $F_X$ in the
previous models, the spectrum in this model is standard in the
sense that there are no exotic chiral fermions, and the quantum
numbers of the matter are typical.  We claim that the superheavy
Higgses ${\bf 10}_H$ and $\overline{\bf 10}_H$ come from a
vector-like pair on the ${\bf 16}^{(d_1)}$ curve, however again it
is not obvious and we are not able to fix the number of such
pairs. In addition, there exist a few exotic $\bf 5$ fields from
the $\bf 10$ curves.

On the other hand, the restrictions of the flux $F_X$ on the
curves in Model 2 are non-vanishing, thus they contribute to the
chirality on the curves. From the information in Table \ref{2-2FX}
we can interpret the matter content to fit the flipped $SU(5)$ GUT
spectrum in Table \ref{22-G-Fsu5}.

\begin{table}[h]
\center
\begin{tabular}{|c|c|c|} \hline
Matter & Rep. & Generation \\ \hline

${\bf 10}_M$ & ${\bf 10}_{1,-1}$ & 3 \\
$\bar{\bf 5}_M$ & $\bar{\bf 5}_{1,3}$ & 3 \\
${\bf 1}_M$ & ${\bf 1}_{1,-5}$ & 3  \\ \hline

${\bf 10}_H+\overline{\bf 10}_H$ & ${\bf 10}_{1,-1}+\overline{\bf 10}_{1,1}$ & 1 \\
\hline

${\bf 5}_h$ & ${\bf 5}_{-2,2}$ & 1 \\
$\bar{\bf 5}_h$ & $\bar{\bf 5}_{-2,-2}$ & 1 \\  \hline \hline

$\bar{\bf 5}$ & $\bar{\bf 5}_{-1,3}$ & 1 \\
${\bf 1}$ &  ${\bf 1}_{-1,5}$ & 1 \\
${\bf 1}$ &  ${\bf 1}_{1,-5}$ & 2 \\ \hline

%\multirow{3}{*}{${\bf 5}+\bar{\bf 5}$ exotics} & ${\bf
%5}_{-2,2}+\bar{\bf 5}_{-2,-2}$ & 4, 3 \\

%& ${\bf 5}_{0,2}+\bar{\bf 5}_{0,-2}$  & -%3,5  \\
%& ${\bf 5}_{2,2}+\bar{\bf 5}_{2,-2}$  & -%0,-1

\multicolumn{3}{|c|}{${\bf 5}+\bar{\bf 5}$ exotics from the $\bf 10$ curves\footnotemark[2]} \\
\hline
\end{tabular}
\caption{Flipped $SU(5)$ spectrum of Model 2 of the (2,2)
factorization in $dP_7$.} \label{22-G-Fsu5}
\end{table}
\footnotetext[2]{The $({\bf 5},\bar{\bf 5})$ exotics from the $\bf
10$ curves of $SO(10)$ can be obtained from Table \ref{2-2FX}. }

In this case, the Yukawa couplings for flipped $SU(5)$ are the
same:
\begin{eqnarray}
\mathcal{W}&\supset&{\bf 10}_{-1,-1M} {\bf 10}_{-1,-1M} {\bf
5}_{2,2h} + {\bf 10}_{-1,-1M} \bar{\bf 5}_{1,3M}\bar{\bf
5}_{0,-2h'} +
\bar{\bf 5}_{1,3M} {\bf 1}_{-1,-5M} {\bf 5}_{0,2h'} \nonumber \\
&+&  {\bf 10}_{-1,-1H} {\bf 10}_{-1,-1H} {\bf 5}_{2,2h} +
\overline{\bf 10}_{1,1H} \overline{\bf 10}_{1,1H} \bar{\bf
5}_{-2,-2h}+\dots.
\end{eqnarray}
The ${\bf 10}+\overline{\bf 10}$ superheavey Higgses are
identified as a vector-like pair from the $\bf 16$ curve. In this
model there are a few unavoidable exotic fields descended from
both $\bf 16$ and $\bf 10$ curves. }

\subsubsection{The Singlet Higgs}

In the flipped $SU(5)$ model, the matter singlet is the
right-handed electron, while it is the right-handed neutrino in
the Georgi-Glashow $SU(5)$ GUT. Different from the $SU(5)$
spectral cover construction, the flipped $SU(5)$ matter singlet is
naturally embedded into the $\bf 16$ representation of $SO(10)$ in
the $SU(4)$ spectral cover configuration.  Thus there is no need
of additional effort to identify it in the spectrum.

Moreover, in flipped $SU(5)$ models, a Yukawa coupling needed to
explain neutrino masses with the seesaw mechanism is
\cite{Nanopoulos:2002qk, Georgi:1979dq}
\begin{equation}
{\bf 10}_{1M}\overline{\bf 10}_{-1H} {\bf 1}_{0\phi}.
\end{equation}
This singlet ${\bf 1}_{0}$ is an $SO(10)$ object and descends
neither from the $\bf 16$ nor from the $\bf 10$ curves. Naively,
one might think that it can be captured by the spectral cover
associated to the adjoint representation in $SU(4)$ and the matter
curve corresponds to $\pm(\lambda_i-\lambda_j)=0$ with $i\neq j$.
The locus would then be given by ~\cite{Caltech:global03}
\begin{equation}
b_0^5\prod_{i<j}^4 (\lambda_i -\lambda_j)^2= -4b_2^3 b_3^2 -27 b_0
b_3^4 +16 b_2^4 b_4 +144 b_0 b_2 b_3^2 b_4 -128 b_0 b_2^2
b_4^2 +256 b_0^2 b_4^3=0. \nonumber\\
\end{equation}
However, this is not the case. In fact, this singlet matter curve
lives in the base $B_3$ instead of the surface $S$ and can not be
described by the spectral cover. To calculate the matter chirality
on this singlet matter curve, we need the information of global
geometry transverse to the surface $S$. In other words, we need to
go beyond the spectral cover construction\footnote{Recently this
singlet has been discussed in\cite{Grimm:2010ez} for the $SU(5)$
GUT, and it is possible to apply the same idea in this case.  We
leave this topic for our future work.}. In the future, we hope
there will be a global understanding of this singlet curve
\cite{Caltech:global03}. Therefore, we just assume this singlet
exists and can provide the above Yukawa coupling.

\section{Conclusions}

In this paper we built flipped $SU(5)$ models from the $SO(10)$
singularity by the $SU(4)$ spectral cover construction in
F-theory.  The ${\bf 10}$ curve in the $SU(4)$ spectral cover
configuration forms a double curve, and there are codimension two
singularities on this curve \cite{Other:global}. It has been also
shown that the net chirality on the {\bf 10} curve vanishes
\cite{Other:global}. In order to obtain more degrees of freedom
and non-zero generation number on the ${\bf 10}$ curve, we split
the $SU(4)$ cover into two factorizations. In the (3,1)
factorization there are two {\bf 16} curves and two ${\bf 10}$
curves on $S$, while in the (2,2) factorization there are two
${\bf 16}$ curves and three ${\bf 10}$ curves.  The fluxes are
also spread over the curves, providing additional parameters for
model building.

We start model building from setting up appropriate $SO(10)$
spectrum on the $\bf 16$ and $\bf 10$ curves. Some Higgs fields,
such as $\bf 210$, $\bf 120$, and ${\bf 126}+\overline{\bf 126}$
breaking the $SO(10)$ gauge group are absent in this construction.
Therefore, we introduce a $U(1)_X$ flux to break $SO(10)$ to
$SU(5)\times U(1)_X$. We interpret the resulting spectrum as a
flipped $SU(5)$ model. The flux may have non-vanishing
restrictions on the curves such that the corresponding chiralities
may be modified. The superheavy Higgs fields ${\bf 10}_H$ and
$\overline{\bf 10}_H$ needed for breaking the gauge group to the
MSSM are not obvious from the spectrum. We assume that they are a
vector-like pair from the ${\bf 16}$ curve including the fermion
representations, but we are not able to fix the number of
such~pairs.

In the (3,1) factorization, we discuss first the construction on
the geometry of the Calabi-Yau fourfold with an embedded $dP_2$
surface constructed in \cite{Caltech:global01}. We demonstrated
three examples. Two of them have three-generation, minimal
$SO(10)$ GUT matter spectra. The $U(1)_X$ flux has always
non-vanishing restrictions on the ${\bf 16}$ curves, while it
generically has vanishing restrictions on the ${\bf 10}$ curves.
Therefore, on a $\bf 16$ curve, the chiralities of the $\bf 10$,
$\bf 5$, and $\bf 1$ representations are modified in the factor of
the $U(1)_X$ charges, and the model no longer has three
generations after the $SO(10)$ gauge symmetry is broken. To solve
this problem, we constructed a four-generation model such that its
corresponding flipped $SU(5)$ spectrum can possess at least three
generations after the $U(1)_X$ flux is turned on. On the other
hand, the $U(1)_X$ flux in the case of $dP_7$ geometry background
\cite{Blumenhagen:global02} can be tuned to have trivial
restrictions on the $\bf 16$ curves so the chiralities remain
untouched.  We presented one three-generation example of the (3,1)
factorization based on this geometry.

In the (2,2) factorization, to have more degrees of freedom for
model building, we focused only on the geometry of the Calabi-Yau
fourfold with an embedded $dP_7$ surface
\cite{Blumenhagen:global02} and presented two examples. The first
was a three-generation flipped $SU(5)$ model from the $SO(10)$
gauge group broken by the flux with trivial restrictions on all
the matter curves. The second example, however, starts from a
four-generation $SO(10)$ model whose gauge group is broken to
$SU(5)\times U(1)_X$ by the flux with non-trivial restrictions on
the matter curves. The resulting chiralities are modified by the
flux restrictions to achieve the spectrum of a three-generation
flipped $SU(5)$ model. Generically, the flipped $SU(5)$ models
from a four-generation $SO(10)$ setup with non-vanishing flux
restrictions to the $\bf 16$ curves results in exotic fields from
the $\bf 16$ curves.

There remain some interesting directions for future research.
First, we could construct $SO(10)$ singularities directly on
Calabi-Yau fourfolds. Some examples in toric geometry are
discussed in \cite{Chen:2010ts}, and it would be interesting to
consider more general fourfolds. Second, the $SO(10)$ singlet is
important for the neutrino mass problem in the flipped $SU(5)$
phenomenology, however the mechanism of defining this singlet
remains unclear. Third, we could investigate flipped $SU(5)$
models that do not descend from a $D_5$ singularity. The flipped
$SU(5)$ models can be built from the anomaly-cancellation of the
$U(1)$s of the monodromy group \cite{Dudas:2010zb} in the
well-studied $SU(5)$ spectral cover configuration in F-theory. A
recent study on the abelian gauge factor from a certain global
restriction of the Tate model \cite{Grimm:2010ez} may be useful to
study the $U(1)$ gauge groups. In addition, it is also exciting if
we can turn on a non-abelian flux to break the $SO(10)$ gauge
symmetry down to a standard-like model, such as the Pati-Salam
model. We leave these questions for our future study.

\renewcommand{\thesection}{}
\section{\hspace{-1cm} Acknowledgments}

We would like to thank A. Braun, J. Knapp, M. Kreuzer, C.
Mayrhofer and J. Marsano for valuable communications and
discussions. We also gratefully acknowledges hospitality and
support from the Simons Center for Geometry and Physics, Stony
Brook. YCC would like to thank the Physics Department at
University of Cincinnati for hospitality during the preparation of
this work. The work of CMC is supported in part by the Austrian
Research Funds FWF under grant I192. The work of YCC is supported
in part by the NSF under grant PHY-0555575 and by Texas A\&M
University.

%\appendix{\bf\Large \hspace{-.85cm} Appendix}

\renewcommand{\theequation}{\thesection.\arabic{equation}}
\setcounter{equation}{0}

%%%%%%%%%%%%%%%%%%%%%%%%%%%%%%%%%%%%%%%%%%%%%%%%%%%%%%%%%%%%%%%%%%%%%%%%%%%%%%%%%%%%%%%%%%%%%%%%%%%%%%%%%%%%%%%%%%

\end{document}